\documentclass[11pt]{article}
\usepackage[reqno]{amsmath}
\usepackage{amssymb}
\usepackage{epsfig}
\usepackage{array}
\usepackage{float}
\usepackage{bbm}
\def\gs{\mathrel{
   \rlap{\raise 0.511ex \hbox{$>$}}{\lower 0.511ex \hbox{$\sim$}}}}
\def\ls{\mathrel{
   \rlap{\raise 0.511ex \hbox{$<$}}{\lower 0.511ex \hbox{$\sim$}}}}

\newcommand{\ba}{\begin{array}{c}}
\newcommand{\baz}{\begin{array}{cc}}
\newcommand{\bad}{\begin{array}{ccc}}
\newcommand{\bav}{\begin{array}{cccc}}
\newcommand{\bea}{\begin{equation} \begin{array}{c}}
\newcommand{\eea}{ \end{array} \end{equation}}
\newcommand{\ea}{\end{array}}
\newcommand{\D}{\displaystyle}
\newcommand{\dms}{\mbox{$\Delta m^2_{\odot}$}}
\newcommand{\dma}{\mbox{$\Delta m^2_{\rm A}$}}
\newcommand{\dmsol}{\mbox{$\Delta m^2_{\odot}$}}
\def\gtap{\mathrel{ \rlap{\raise 0.511ex \hbox{$>$}}{\lower 0.511ex
   \hbox{$\sim$}}}} 
\def\ltap{\mathrel{ \rlap{\raise 0.511ex
   \hbox{$<$}}{\lower 0.511ex \hbox{$\sim$}}}}
\newcommand{\deltaatm}{\mbox{$\Delta m^2_{31}$}}
\newcommand{\deltasol}{\mbox{$ \Delta m^2_{21}$}}

\newcommand{\betabeta}{\mbox{$(\beta \beta)_{0 \nu}$}}

\newcommand{\meff}{\mbox{$\left|\langle m\rangle\right|$}}

\newcommand{\hbeta}{$\mbox{}^3 {\rm H}$ $\beta$-decay }

\newcommand{\pmns}{\mbox{$U_{\rm PMNS}$}}

\def\ie{\hbox{\it i.e.}{}}
\def\eg{\hbox{\it e.g.}{}}
\def\etc{\hbox{\it etc}{}}
\renewcommand{\thefootnote}{\fnsymbol{footnote}}
\begin{document}
\begin{titlepage}
\hfill
\vbox{
  \halign{#\hfil  \cr
    SISSA 78/2005/EP \cr
    IC/2005/102 \cr
    TUM-HEP-606/05 \cr
    hep-ph/0510404 \cr}}
\vspace*{4mm}
\begin{center}
{\bf \large{The See-Saw Mechanism, Neutrino Yukawa Couplings, \\
LFV Decays $l_i \to l_j +\gamma$ and Leptogenesis}} \\
\vspace*{7mm}
{\ S.~T.~Petcov ${}^{a, b)}$}\footnote[1]{Also at: Institute
of Nuclear Research and Nuclear Energy, Bulgarian Academy %
of Sciences, 1784 Sofia, %
Bulgaria},%
{\ W.~Rodejohann${}^{c)}$},%
{\ T.~Shindou${}^{a,b)}$}\footnote[2]{E-mail: shindou@sissa.it},%
{\ Y.~Takanishi${}^{d)}$}\footnote[3]{Address after 22 November 2005:
Scuola Internazionale Superiore di Studi
Avanzati, I-34014 Trieste, Italy,
E-mail: yasutaka@sissa.it}

\vspace*{0.7cm}

{${}^{a)}${\it Scuola Internazionale Superiore di Studi
Avanzati, I-34014 Trieste, Italy}
\vskip .3cm

${}^{b)}${\it Istituto Nazionale di Fisica Nucleare,
Sezione di Trieste, I-34014 Trieste, Italy}
\vskip .3cm

${}^{c)}${\it Physik-Department, Technische Universit\"at M\"unchen, \\ 
James-Franck-Stra{\ss}e, D-85748 Garching, Germany}

\vskip .3cm

${}^{d)}${\it The Abdus Salam International Centre for Theoretical
Physics, \\ Strada Costiera 11, I-34100 Trieste, Italy}} \\

%\vspace*{1cm}
\vskip 1cm
\end{center}
\begin{abstract}
%%%%%%%%%%%%%%%%%%%%%%%%%%%%%%%%%%%%%%%%%%%%%%%%%%%%%%%
  \indent\  The LFV charged lepton decays $\mu \to e + \gamma$, $\tau
  \to e + \gamma$ and $\tau \to \mu + \gamma$ and thermal leptogenesis
  are analysed in the MSSM with see-saw mechanism of neutrino mass
  generation and soft SUSY breaking with universal boundary
  conditions. The case of hierarchical heavy Majorana neutrino mass
  spectrum, $M_1 \ll M_2 \ll M_3$, is investigated.  Leptogenesis
  requires $M_1 \gtap 10^{9}$ GeV.  Considering the natural range of
  values of the heaviest right-handed Majorana neutrino mass, $M_3
  \gtap 5\times10^{13}$ GeV, and assuming that the soft SUSY breaking
  universal gaugino and/or scalar masses have values in the range of
  ${\rm few}\times 100$ GeV, we derive the combined constraints, which
  the existing stringent upper limit on the $\mu \to e + \gamma$ decay
  rate and the requirement of successful thermal leptogenesis impose
  on the neutrino Yukawa couplings, heavy Majorana neutrino masses and
  SUSY parameters.  Results for the three possible types of light
  neutrino mass spectrum -- normal and inverted hierarchical and
  quasi-degenerate -- are obtained.
%%%%%%%%%%%%%%%%%%%%%%%%%%%%%%%%%%%%%%%%%%%%%%%%%%%%%%%
%\vskip 5.5mm \noindent\
%PACS numbers: \\
%\vskip -3mm \noindent\
%Keywords: \\
\end{abstract}
\vskip 4.5cm
October 2005
\vskip .5cm
\end{titlepage}

%%%%%%%%%%%%%%%%%%%%%%%%%%%%%%%%%%%%%%%%%%%%%%%%%%%%%%%
\newpage
\renewcommand{\thefootnote}{\arabic{footnote}}
\setcounter{footnote}{0}
\setcounter{page}{1}
\section{\label{sec:intro}Introduction}
\indent\ The experiments with solar, atmospheric, reactor and
accelerator neutrinos~\cite{sol,SKsolaratm,SNO123,KamLAND,K2K} have
provided during the last several years compelling evidence for the
existence of non-trivial 3-neutrino mixing in the weak charged-lepton
current (see, $\eg$,~\cite{STPNu04}):
%%%%%%%%%%%%%%%%%%%
\begin{equation}
\nu_{l \mathrm{L}}  = \sum_{j=1}^{3} U_{l j} \, \nu_{j \mathrm{L}},~~
l  = e,\mu,\tau,
\label{3numix}
\end{equation}
%%%%%%%%%%%%%%%%%%%
%
\noindent where 
$\nu_{lL}$ are the flavour neutrino fields, $\nu_{j \mathrm{L}}$ is
the field of neutrino $\nu_j$ having a mass $m_j$ and $U$ is the
Pontecorvo-Maki-Nakagawa-Sakata (PMNS) mixing
matrix~\cite{BPont57}, $U \equiv \pmns$.  The existing data, including
the data from the \hbeta experiments~\cite{MoscowH3Mainz} imply that
the massive neutrinos $\nu_j$ are significantly lighter than the
charged leptons and quarks: $m_{j} < 2.3$ eV (95\%
C.L.)~\footnote{More stringent upper limit on $m_j$ follows from the
  constraints on the sum of neutrino masses obtained from
  cosmological/astrophysical observations, namely, the CMB data of the
  WMAP experiment combined with data from large scale structure
  surveys (2dFGRS, SDSS)~\cite{WMAPnu}: $\sum_{j} m_j < (0.7 - 2.0)$
  eV (95\% C.L.), where we have included a conservative estimate of
  the uncertainty in the upper limit (see, $\eg$,~\cite{Hanne03}).}.
A natural explanation of the smallness of neutrino masses is provided
by the see-saw mechanism of neutrino mass generation~\cite{seesaw}.
The see-saw mechanism predicts the light massive neutrinos $\nu_j$ to
be Majorana particles. An integral part of the mechanism are the heavy
right-handed (RH) Majorana neutrinos~\cite{Pont67}.  In grand unified
theories (GUT) the masses of the heavy RH Majorana neutrinos are
typically by a few to several orders of magnitude smaller than the
scale of unification of the electroweak and strong interactions,
$M_{\rm GUT} \cong 2\times 10^{16}$ GeV.  In this case the
CP-violating decays of the heavy RH Majorana neutrinos in the Early
Universe could generate, through the leptogenesis scenario, the
observed baryon asymmetry of the Universe~\cite{LeptoG}.

The existence of the flavour neutrino mixing, eq.~(\ref{3numix}),
implies that the individual lepton charges, $L_l$, $l =e,\mu,\tau$,
are not conserved (see, $\eg$,~\cite{BiPet87}), and processes like
$\mu^- \rightarrow e^- + \gamma$, $\mu^{-} \rightarrow e^{-} + e^{+} +
e^{-}$, $\tau^- \rightarrow e^- + \gamma$, $\tau^- \rightarrow \mu^- +
\gamma$, $\mu^{-} + (A,Z) \rightarrow e^{-} + (A,Z)$, $\etc$.  should
take place.  Stringent experimental upper limits on the branching
ratios and relative cross sections of the indicated $|\Delta L_l| = 1$
decays and reactions have been obtained~\cite{mega,PDG04,BaBar05} (90\% C.L.):
%%%%%%%%%%%%%%%%%%%%%%%%%%%%%%%%%%%%%%%
\begin{equation}
\ba
\text{B}(\mu \to e+\gamma) < 1.2\times 10^{-11},~~%\\[0.24cm]
\text{B}(\mu \to 3e) < 1.2\times 10^{-12},~~
\text{R}(\mu^{-} + \text{Ti} \rightarrow e^{-} + \text{Ti}) <
4.3\times 10^{-12}\;, \\[0.24cm]
\text{B}(\tau \to \mu +\gamma) < 6.8\times 10^{-8}~,~~~%\\[0.24cm]
\text{B}(\tau \to e+ \gamma) < 1.1\times 10^{-7}~.   
\ea 
\end{equation}
%%%%%%%%%%%%%%%%%%%%%%%%%%%%%%%%%%%%%%%%
%
%\vspace{-0.2cm}
\noindent Future experiments with increased 
sensitivity can reduce the current bounds on $\text{B}(\mu\to
e+\gamma)$, $\text{B}(\tau\to \mu +\gamma)$ and on $\text{R}(\mu^{-} +
(A,Z) \rightarrow e^{-} + (A,Z))$ by a few orders of magnitude 
(see, $\eg$,~\cite{Kuno99}).  In the experiment MEG under preparation at
PSI~\cite{psi} it is planned to reach a sensitivity to
%%%%%%%%%%%%%%%%%%%%%%%%%%%%%
\begin{equation}
 \text{B}(\mu \to e+\gamma)\sim (10^{-13} - 10^{-14})\,. 
\end{equation}
%%%%%%%%%%%%%%%%%%%%%%%%%%%%%
%
\indent\ It has been noticed a long time ago that in SUSY (GUT)
theories with see-saw mechanism of neutrino mass generation, the rates
and cross sections of the LFV processes can be strongly
enhanced~\cite{BorzMas86}.  If the SUSY breaking occurs via soft terms
with universal boundary conditions at a scale $M_X$ above the RH
Majorana neutrino mass scale $M_R$, $M_{X}>M_R$~\footnote{The
  possibility of ``flavour-blind'' SUSY breaking of interest is
  realised, $\eg$, in gravity-mediation SUSY breaking scenarios (see,
  $\eg$,~\cite{GMFB}).}, the renormalisation group (RG) effects
transmit the LFV from the neutrino mixing at $M_{X}$ to the effective
mass terms of the scalar leptons at $M_R$ even if the soft SUSY
breaking terms at $M_X$ are flavour symmetric and conserve the lepton
charges $L_l$.  As a consequence of these RG-induced new LFV terms in
the effective Lagrangian at $M_R < M_X$, the LFV processes can proceed
with rates and cross sections which are within the sensitivity of
presently operating and future planned 
experiments~\cite{BorzMas86,Hisano96} 
(see also, $\eg$,~\cite{Iba01,JohnE,Saclay0105,PPY03,PPR3,PPTY03,Eiichi05,PShinYasu05}).
In contrast, in the non-supersymmetric case, the rates and cross
sections of the LFV processes are suppressed by the factor~\cite{SP76}
(see also~\cite{BPP77}) $(m_{j}/M_W)^4 < 6.7\times 10^{-43}$, $M_W$
being the $W^{\pm}$ mass, which renders them unobservable.

One of the basic ingredients of the see-saw mechanism is the matrix of
neutrino Yukawa couplings, $\mathbf{Y_{\nu}}$.  Leptogenesis depends
on $\mathbf{Y_{\nu}}$ as well~\cite{LeptoG} (see
also~\cite{LGBDiBP05,CERN04} and the references quoted therein).  In
the large class of SUSY models with see-saw mechanism and SUSY
breaking mediated by flavour-universal soft terms at a scale
$M_{X}>M_R$ we will consider, the probabilities of LFV processes also
depend strongly on $\mathbf{Y_{\nu}}$ (see,
$\eg$,~\cite{Iba01,JohnE}).  The matrix $\mathbf{Y_{\nu}}$ can be
expressed in terms of the light neutrino and heavy RH neutrino masses,
the neutrino mixing matrix $\pmns$, and an orthogonal matrix
$\mathbf{R}$~\cite{Iba01}.  Leptogenesis can take place only if
$\mathbf{R}$ is complex.  The matrix $\mathbf{Y_{\nu}}$ depends, in
particular, on the Majorana CP-violation (CPV) phases in the PMNS
matrix $\pmns$~\cite{BHP80}~\footnote{Obtaining information about the
  Majorana CPV phases in the PMNS matrix $\pmns$ if the massive
  neutrinos are proved to be Majorana particles would be a remarkably
  challenging problem. The oscillations of flavour neutrinos, $\nu_{l}
  \rightarrow \nu_{l'}$ and $\bar{\nu}_{l} \rightarrow
  \bar{\nu}_{l'}$, $l,l'=e,\mu,\tau$, are insensitive to the two
  Majorana phases in $\pmns$~\cite{BHP80,Lang87}. The only feasible
  experiments that at present have the potential of establishing the
  Majorana nature of light neutrinos $\nu_j$ and of providing
  information on the Majorana phases in $\pmns$ are the experiments
  searching for neutrinoless double beta ($\betabeta$)-decay, $(A,Z)
  \rightarrow (A,Z+2) + e^- + e^-$ (see,
  $\eg$,~\cite{BiPet87,APSbb0nu,BPP1,STPFocusNu04}).  }.  It was shown
in~\cite{PPY03,PShinYasu05} that if the heavy Majorana neutrinos are
quasi-degenerate in mass, the Majorana phases can affect significantly
the predictions for the rates of LFV decays $\mu \rightarrow e +
\gamma$, $\tau \rightarrow e + \gamma$, $\etc$.  in the class of SUSY
theories of interest.

The matrix $\mathbf{Y_{\nu}}$ can be defined, strictly speaking, only
at scales not smaller than $M_R$. The probabilities of LFV processes
depend on $\mathbf{Y_{\nu}}$ at the scale $M_R$, $\mathbf{Y_{\nu}}
=\mathbf{Y_{\nu}}(M_R)$.  In order to evaluate $\mathbf{Y_{\nu}}(M_R)$
one has to know, in general, the light neutrino masses $m_j$ and the
mixing matrix $\pmns$ at $M_R$, $\ie$, one has to take into account
the renormalisation group (RG) ``running'' of $m_j$ and $\pmns$ from
the scale $M_Z \sim 100$ GeV, at which the neutrino mixing parameters
are measured, to the scale $M_R$ 
(see, $\eg$,~\cite{RGrunU,PShinYasu05} and the references quoted therein).
However, if the RG running of $m_j$ and $\pmns$ is sufficiently small,
$\mathbf{Y_{\nu}}(M_R)$ will depend on the values of the light
neutrino masses $m_j$ and the mixing angles and CP-violation phases in
$\pmns$ at the scale $M_Z$.

Working in the framework of the class of SUSY theories with see-saw
mechanism and soft SUSY breaking with flavour-universal boundary
conditions at a scale $M_X>M_R$, we investigate in the present article
the combined constraints, which the existing stringent upper limit of
the $\mu \to e + \gamma$ decay rate and the requirement of successful
thermal leptogenesis impose on the neutrino Yukawa couplings, heavy
Majorana neutrino masses and on the SUSY parameters.  The case of
hierarchical heavy Majorana neutrino mass spectrum, $M_1 \ll M_2 \ll
M_3$, is considered.  Leptogenesis requires $M_1 \gtap 10^{9}$ GeV.
The analysis is performed assuming that the heaviest RH Majorana
neutrino has a mass $M_3 \gtap 5\times10^{13}$ GeV, and that the soft
SUSY breaking universal gaugino and/or scalar masses (at the scale
$M_X$) have values in the range of ${\rm few}\times 100$ GeV.  One
typically gets $M_3 \gtap 5\times10^{13}$ GeV in SUSY GUT theories
with see-saw mechanism of neutrino mass generation (see,
$\eg$,~\cite{GUTM3}).  If the SUSY breaking universal gaugino and/or
scalar masses have values in the ${\rm few}\times 100$ GeV range,
supersymmetric particles will be observable in the experiments under
preparation at the LHC (see, $\eg$,~\cite{LHCSUSY}).  We find that
under the indicated assumptions, the existing stringent upper limit on
the $\mu \to e + \gamma$ decay rate cannot be satisfied, unless the
terms proportional to $M_3$ in the $\mu \to e + \gamma$ decay
amplitude are absent or strongly suppressed.  The requisite
``decoupling'' of the terms $\propto M_3$ from the $\mu \to e +
\gamma$ decay amplitude is realised if the matrix $\mathbf{R}$ has a
specific form which admits a parametrisation with just one complex
angle.  Using the latter we obtain results for the three types of
light neutrino mass spectrum -- normal and inverted hierarchical (NH
and IH), and quasi-degenerate (QD).  For each of the three types of
spectrum we derive the leptogenesis lower bound on the mass of the
lightest RH Majorana neutrino $M_1$.  The lower bounds thus found in
the cases of IH and QD spectrum are $\sim 10^{13}$ GeV.  The upper
limit on $\text{B}(\mu \to e+\gamma)$ in these two cases can be
satisfied for specific ranges of values of the soft SUSY breaking
parameters implying relatively large masses of the supersymmetric
particles.  Using these soft SUSY breaking parameters we derive
predictions for $\text{B}(\mu \to e+\gamma)$, $\text{B}(\tau \to
e+\gamma)$ and $\text{B}(\tau \to \mu +\gamma)$ which are compatible
with the requirement of successful leptogenesis.

Our analysis is performed under the condition of negligible RG effects
for the light neutrino masses $m_j$ and the mixing angles and
CP-violation phases in $\pmns$.  The RG effects in question are
negligible in the class of SUSY theories we are considering in the
case of hierarchical light neutrino mass spectrum 
(see, $\eg$,~\cite{RGrunU,PShinYasu05}).  The same is valid for quasi-degenerate
$\nu_j$ mass spectrum provided the parameter $\tan\beta < 10$,
$\tan\beta$ being the ratio of the vacuum expectation values of the
up- and down-type Higgs doublet fields in SUSY extensions of the
Standard Theory.

%%%%%%%%%%%%%%%%%%%%%%%%%%%%%%%%%%%%%%%%%%%%%%%
%
\section{\large{General Considerations}}
%
%%%%%%%%%%%%%%%%%%%%%%%%%%%%%%%%%%%%%%%%%%%%%%%%
%%%%%%%%%%%%%%%%%%%%%%%%%%%%%%%%%%%%%%%%%%%%%%
%
\subsection{\large{Neutrino Mixing Parameters from Neutrino Oscillation Data}}
%
%%%%%%%%%%%%%%%%%%%%%%%%%%%%%%%%%%%%%%%%%%%%%
%
\indent We will use the standard parametrisation of the
PMNS matrix $\pmns$ (see, $\eg$,~\cite{BPP1}): 
%%%%%%%%%%%%%%%%%%%%%%%%%%%%%%%%%%%
\bea 
\label{eq:Upara}
\pmns = \left( \bad 
 c_{12} c_{13} & s_{12} c_{13} & s_{13} e^{-i \delta} \\[0.2cm]   
 -s_{12} c_{23} - c_{12} s_{23} s_{13} e^{i \delta} 
 & c_{12} c_{23} - s_{12} s_{23} s_{13} e^{i \delta} & s_{23} c_{13} \\[0.2cm] 
 s_{12} s_{23} - c_{12} c_{23} s_{13} e^{i \delta} & 
 - c_{12} s_{23} - s_{12} c_{23} s_{13} e^{i \delta} & c_{23} c_{13} \\ 
\ea   \right) 
{\rm diag}(1, e^{i \frac{\alpha}{2}}, e^{i \frac{\beta_M}{2}}) \, ,
\eea
%%%%%%%%%%%%%%%%%%%%%%%%%%%%%%%%%
%
\noindent where 
$c_{ij} = \cos\theta_{ij}$, $s_{ij} = \sin\theta_{ij}$, the angles
$\theta_{ij} = [0,\pi/2]$, $\delta = [0,2\pi]$ is the Dirac
CP-violating phase and $\alpha$ and $\beta_M$ are two Majorana
CP-violation phases~\cite{BHP80,SchValle80D81}.  One can identify the
neutrino mass squared difference responsible for solar neutrino
oscillations, $\dms$, with $\Delta m^2_{21} \equiv m^2_2 - m^2_1$,
$\dms = \Delta m^2_{21} > 0$.  The neutrino mass squared difference
driving the dominant $\nu_{\mu} \rightarrow \nu_{\tau}$
($\bar{\nu}_{\mu} \rightarrow \bar{\nu}_{\tau}$) oscillations of
atmospheric $\nu_{\mu}$ ($\bar{\nu}_{\mu}$) is then given by
$\left|\dma\right|=\left|\Delta m^2_{31}\right|\cong \left|\Delta m^2_{32}\right| \gg \Delta m^2_{21}$.
The corresponding solar and atmospheric neutrino mixing angles,
$\theta_{\odot}$ and $\theta_{\rm A}$, coincide with $\theta_{12}$ and
$\theta_{23}$, respectively.  The angle $\theta_{13}$ is limited by
the data from the CHOOZ and Palo Verde experiments~\cite{CHOOZPV}.

The existing neutrino oscillation data allow us to determine $\Delta
m^2_{21}$, $\left|\Delta m^2_{31}\right|$, $\sin^2\theta_{12}$ and
$\sin^22\theta_{23}$ with a relatively good precision and to obtain
rather stringent limits on $\sin^2\theta_{13}$ (see,
$\eg$,~\cite{SKsolaratm,BCGPRKL2,3nuGlobal}).  The best fit values of
$\Delta m^2_{21}$, $\sin^2\theta_{12}$, $\left|\Delta m^2_{31}\right|$ and
$\sin^22\theta_{23}$ read~\footnote{The data imply, in particular,
  that maximal solar neutrino mixing is ruled out at $\sim 6\sigma$;
  at 95\% C.L.\ one finds $\cos 2\theta_\odot \geq
  0.26$~\cite{BCGPRKL2}, which has important
  implications~\cite{PPSNO2bb}.}:
%%%%%%%%%%%%%%%%%%%%%%%%%%%%%%%%%%%
\begin{equation}
\label{bfvsol}
\ba
\deltasol = 8.0\times 10^{-5}~{\rm eV^2},~~
\sin^2\theta_{12} = 0.31~, % \\[0.25cm]
\ea
\end{equation}
%%%%%%%%%%%%%%%%%%%%%%%%%%%%%%%%%%%%
% \vspace{-0.5cm}
%%%%%%%%%%%%%%%%%%%%%%%%%%%%%%%%% 
\begin{equation}
\label{eq:atmrange}
\ba
\left|\deltaatm\right| = 2.1\times 10^{-3}~{\rm eV^2}~,~~\sin^22\theta_{23} = 1.0
~, %\\  [0.25cm]
\ea
\end{equation}
%%%%%%%%%%%%%%%%%%%%%%%%%%%%%%%%%
% 
\noindent
A combined 3-$\nu$ oscillation
analysis of the solar neutrino, 
KamLAND and CHOOZ data gives~\cite{BCGPRKL2}
%%%%%%%%%%%%%%%%%%%%%%%%%%%%%%
\begin{equation}
\sin^2\theta_{13} < 0.024~(0.044),~~~~\mbox{at}~95\%~(99.73\%)~{\rm C.L.}
\label{th13}
\end{equation}
%%%%%%%%%%%%%%%%%%%%%%%%%%%%%%%
%
 The neutrino oscillation parameters
$\deltasol$, $\sin^2\theta_{12}$,
$|\deltaatm|$ and $\sin^22\theta_{23}$
are determined by the existing
data at 3$\sigma$
with an error of approximately 12\%, 24\%,
50\% and 16\%, respectively.
These parameters can (and very
likely will) be measured with much
higher accuracy in the future (see,
$\eg$,~\cite{STPNu04}). In all further numerical 
estimates we use the best fit values of 
$\deltasol$, $\sin^2\theta_{12}$,
$|\deltaatm|$ and $\sin^22\theta_{23}$. 
Whenever the parameter $\sin\theta_{13}$ 
is also relevant in the calculations, 
we specify the value used.

  The sign of $\dma = \deltaatm $, as it is well known, cannot be
determined from the present (SK atmospheric neutrino and K2K) data.
The two possibilities, $\Delta m^2_{31(32)} > 0$ or $\Delta
m^2_{31(32)} < 0$ correspond to two different
types of $\nu$-mass spectrum:\\
-- {\it with normal hierarchy} % (or ordering),
$m_1 < m_2 < m_3$, $\dma=\Delta m^2_{31} >0$, and \\
-- {\it with inverted hierarchy} % (ordering)
$m_3 < m_1 < m_2$, $\dma =\Delta m^2_{32}< 0$. \\
\noindent Depending on the sign of \dma, ${\rm sgn}(\dma)$, and 
the value of the lightest neutrino mass,
${\rm min}(m_j)$, the $\nu$-mass  spectrum can be\\
-- {\it Normal Hierarchical}: $m_1 \ll m_2 \ll m_3$,
$m_2 \cong (\dmsol)^ {\frac{1}{2}} \sim$ 0.009 eV,
$m_3 \cong \left|\dma\right|^{\frac{1}{2}} \sim$ 0.045 eV;\\
-- {\it Inverted Hierarchical}: $m_3 \ll m_1 < m_2$,
with $m_{1,2} \cong \left|\dma\right|^{\frac{1}{2}}\sim$ 0.045 eV; \\
-- {\it Quasi-Degenerate (QD)}: $m_1 \cong m_2 \cong m_3 \cong m$,
$m_j^2 \gg \left|\dma\right|$, $m\gtap 0.10$~eV.

%%%%%%%%%%%%%%%%%%%%%%%%%%%%%%%%%%%%%%%%%%%%%%%
%
\subsection{\large{The See-Saw Mechanism and Neutrino Yukawa Couplings}}
%
%%%%%%%%%%%%%%%%%%%%%%%%%%%%%%%%%%%%%%%%%%%%%%%%
%
\indent\ We consider the minimal supersymmetric standard model with RH
neutrinos and see-saw mechanism of neutrino mass generation (MSSMRN).
In the framework of MSSMRN one can always choose a basis in which both
the matrix of charged lepton Yukawa couplings, $\mathbf{Y_{\rm E}}$,
and the Majorana mass matrix of the heavy RH neutrinos,
$\mathbf{M_{\rm N}}$, are real and diagonal.  Henceforth, we will work
in that basis and will denote by $\mathbf{D_{\rm N}}$ the
corresponding diagonal RH neutrino mass matrix, $\mathbf{D_{\rm N}} =
{\rm diag}(M_1,M_2,M_3)$, with $M_j > 0$ and $M_1 < M_2 < M_3$. The
largest mass $M_3$ will be standardly assumed to be of the order of,
or smaller than, the GUT scale $M_{\rm GUT} \simeq 2\times 10^{16}$
GeV.

Below the see-saw scale, $M_R = {\rm min}(M_j)$, the heavy RH neutrino
fields $N_j$ are integrated out, and as a result of the electroweak
symmetry breaking, the left-handed (LH) flavour neutrinos acquire a
Majorana mass term:
%%%%%%%%%%%%%%%%%%%%%%%%%%%%%%%%%%%%%%
\begin{align}
\mathcal{L}_{m}^{\nu} = 
- \frac{1}{2}~\bar{\nu}^{C}_{Rj}~(m_{\nu})^{jk}~\nu_{Lk} + h.c.~,
\end{align}
%%%%%%%%%%%%%%%%%%%%%%%%%%%%
%
where $\nu^{C}_{Rj} \equiv C (\bar{\nu}_{Lj})^{T}$ and
%%%%%%%%%%%%%%%%%%%%%%%%%%%
\begin{align}
(m_{\nu})^{ij} = 
v_u^2~(Y_{\nu}^T)^{ik}(M_{\rm N}^{-1})^{kl}(Y_{\nu})^{lj}\;.
\label{mnuKN}
\end{align}
%%%%%%%%%%%%%%%%%%%%%%%%%%%%
% 
Here $v_u = v \sin\beta$, where $v = 174$ GeV and $\tan\beta$ is the
ratio of the vacuum expectation values of up-type and down-type Higgs
fields, and $\mathbf{Y}_{\nu}$ is the matrix of neutrino Yukawa
couplings. The neutrino mass matrix $\mathbf{m}_{\nu}$ is related to
the light neutrino masses $m_j$ and the PMNS mixing matrix as follows
%%%%%%%%%%%%%%%%%%%%%%%%%%%%%%%
\begin{align}
(m_{\nu})^{ij}=(U^*)^{ik}m_k(U^{\dagger})^{kj}\;.
\label{mnuU}
\end{align}
%%%%%%%%%%%%%%%%%%%%%%%%%%%%%%%%
%
Using~(\ref{mnuKN}) and~(\ref{mnuU}), we can rewrite the ``matching
condition'' at the energy scale $M_R$ in the form
%%%%%%%%%%%%%%%%%%%%%%%%%%%%
\begin{align}
  \mathbf{U}^*~\mathbf{D}_{\nu}~ \mathbf{U}^{\dagger} =
  v_u^2~\mathbf{Y}_{\nu}^T \mathbf{M}_{\rm N}^{-1}\mathbf{Y}_{\nu}\;.
\end{align}
%%%%%%%%%%%%%%%%%%%%%%%%%%%%%
%
where $\mathbf{D}_{\nu} = \mathrm{diag}(m_1,m_2,m_3)$.  Thus, in the
basis in which the RH neutrino mass matrix is diagonal
$\mathbf{M}_{\rm N} = \mathbf{D}_{\rm N}$, the matrix of neutrino
Yukawa couplings at $M_R$ can be parametrised as~\cite{Iba01}
%%%%%%%%%%%%%%%%%%%%%%%%%%%%
\begin{align}
  \mathbf{Y}_{\nu}(M_R) = \frac{1}{v_u}
  \sqrt{\mathbf{D}_N}~\mathbf{R}~
  \sqrt{\mathbf{D}_{\nu}}~\mathbf{U}^{\dagger}\;.
\label{eq_para_yn}
\end{align}
%%%%%%%%%%%%%%%%%%%%%%%%%%%
%
Here $\mathbf{R}$ is a complex orthogonal 
matrix~\footnote{Equation~(\ref{eq_para_yn}) represents the so-called ``orthogonal''
  parametrisation of $\mathbf{Y}_{\nu}$.  In certain cases it is more
  convenient to use the ``bi-unitary'' parametrisation~\cite{PPR3}
  $\mathbf{Y}_{\nu} = \mathbf{U}^{\dagger}_{R} \mathbf{Y}^{\rm
    diag}_{\nu}~\mathbf{U}_{L}$, where $\mathbf{U}_{\rm L,R}$ are
  unitary matrices and $\mathbf{Y}^{\rm diag}_{\nu}$ is a real
  diagonal matrix.  The orthogonal parametrisation is better adapted
  for our analysis and we will employ it in what follows.}
$\mathbf{R}^T\mathbf{R}= \mathbf{1}$.

In what follows we will investigate the case when the RG running of
$m_j$ and of the parameters in $\pmns$ from $M_Z$ to $M_R$ is
relatively small and can be neglected. This possibility is realised in
the class of theories under discussion for sufficiently small values
of $\tan\beta$ and/or of the lightest neutrino mass 
${\rm min}(m_j)$~\cite{PShinYasu05}, $\eg$, for $\tan\beta \ltap 10$ and/or 
${\rm  min}(m_j) \ltap 0.05$ eV.  Under the indicated condition
$\mathbf{D}_{\nu}$ and $\mathbf{U}$ in eq.~(\ref{eq_para_yn}) can be
taken at the scale $\sim M_Z$, at which the neutrino mixing parameters
are measured.

As is well-known and we shall discuss further, in the case of soft
SUSY breaking mediated by soft flavour-universal terms at $M_X>M_R$,
the predicted rates of LFV processes such as $\mu\to e + \gamma$ decay
are very sensitive to the off-diagonal elements of
%%%%%%%%%%%%%%%%%%%%%%%%%%%%%%%%%%%%%
\begin{align}
\mathbf{Y}_{\nu}^{\dagger}(M_R)\mathbf{Y}_{\nu}(M_R)
= \frac{1}{v_u^2}~
\mathbf{U}\sqrt{\mathbf{D}_{\nu}}~\mathbf{R}^{\dagger}~
\mathbf{D}_N~\mathbf{R}~\sqrt{\mathbf{D}_{\nu}}
\mathbf{U}^{\dagger}\;,
\label{YnudYnu}
\end{align}
%%%%%%%%%%%%%%%%%%%%%%%%%%%%%%%%%%%%%
%
while leptogenesis depends on~\cite{LeptoG} 
(see also~\cite{LGBDiBP05,CERN04} and the references quoted therein)
%%%%%%%%%%%%%%%%%%%%%%%%%%%%%%%%%%%%%
\begin{align}
\mathbf{Y}_{\nu}(M_R)\mathbf{Y}_{\nu}^{\dagger}(M_R)
= \frac{1}{v_u^2}~
\sqrt{\mathbf{D}_{\rm N}}~\mathbf{R}~
\mathbf{D}_{\nu}~\mathbf{R}^{\dagger}~\sqrt{\mathbf{D}_{\rm N}}\;.
\label{YnuYnud}
\end{align}
%%%%%%%%%%%%%%%%%%%%%%%%%%%%%%%%%%%%%
%
In such a way, the matrix of neutrino Yukawa couplings
$\mathbf{Y_{\nu}}$ connects in the see-saw theories the light neutrino
mass generation with leptogenesis; in SUSY theories with SUSY breaking
mediated by soft flavour-universal terms in the Lagrangian at
$M_X>M_R$, $\mathbf{Y_{\nu}}$ links the light neutrino mass generation
and leptogenesis with LFV processes (see, $\eg$,~\cite{PPY03,PPR3}).

%%%%%%%%%%%%%%%%%%%%%%%%%%%%%%%%%%%%%%%%%%%%%%%
%
\subsection{\large{The LFV Decays $l_i\rightarrow l_j + \gamma$}}
%
%%%%%%%%%%%%%%%%%%%%%%%%%%%%%%%%%%%%%%%%%%%%%%%%
\indent\ As was indicated in the Introduction, in the class of theories we
consider, one of the effects of RG running from $M_X$ to $M_R < M_X$
is the generation of new contributions in the amplitudes of the LFV
processes~\cite{BorzMas86,Hisano96}.  In the ``mass insertion'' and
leading-log approximations (see, $\eg$,~\cite{Hisano96,JohnE,PPTY03}),
the branching ratio of $l_i\to l_j + \gamma$ decay due to the new
contributions has the following form
%%%%%%%%%%%%%%%%%%%%%%%%%%%%%%%%%
\begin{equation}
\text{B}(l_i\to l_j + \gamma)\cong 
\frac{\Gamma(l_i\to e\nu\bar{\nu})}{\Gamma_{\text{total}}(l_i)}
\frac{\alpha_{\text{em}}^3}{G_F^2m_S^8}
\left|\frac{(3 + a_0^2)m_0^2}{8\pi^2}\right|^2
\left|\sum_k \left(\mathbf{Y_{\nu}^{\dagger}}\right)_{ik}~\ln\frac{M_{X}}{M_k}~\left(\mathbf{Y_{\nu}}\right)_{kj}
\right|^2\tan^2\beta\;,
\label{eq_ijg}
\end{equation}
%%%%%%%%%%%%%%%%%%%%%%%%%%%%%%%%%%%%%%%%%%
%
where $i\neq j=1,2,3$, $l_1,l_2,l_3\equiv e,\mu,\tau$,
$m_0$ and $A_0 = a_0m_0$ are the universal
scalar masses and trilinear scalar couplings at 
$M_X$ and $m_S$ represents SUSY particle mass.  It was 
shown in~\cite{PPTY03} that in most of the relevant soft 
SUSY breaking parameter space, the expression
%%%%%%%%%%%%%%%%%%%%%%%%%%%%%%%%%%%%%%%%%%
\begin{align}
m_S^8\simeq 0.5~m_0^2~m_{1/2}^2~(m_0^2 + 0.6 ~m_{1/2}^2)^2\;,
\label{eq_ms}
\end{align}
%%%%%%%%%%%%%%%%%%%%%%%%%%%%%%%%%%%%%%%%%%
%
$m_{1/2}$ being the universal gaugino mass at $M_X$, gives an
excellent approximation to the results obtained in a full
renormalisation group analysis, $\ie$, without using the leading-log
and the mass insertion approximations.  It proves useful to consider
also the ``double'' ratios,
%%%%%%%%%%%%%%%%%%%%%%%%%%%%%%
\begin{align}
  \text{R}(21/31) \equiv \frac{\text{B}(\mu \to e + \gamma)}
  {\text{B}(\tau \to e + \gamma)}~\text{B}(\tau \to
  e\nu_{\tau}\bar{\nu}_e)\;,~ \text{R}(21/32) \equiv
  \frac{\text{B}(\mu \to e + \gamma)} {\text{B}(\tau \to \mu +
    \gamma)} ~\text{B}(\tau \to e\nu_{\tau}\bar{\nu}_e)\;,
\label{DoubleR}
\end{align}
%%%%%%%%%%%%%%%%%%%%%%%%%%%%%%%%%%%%%%%%%%%%%%%
%
which are essentially independent of the SUSY parameters.

  To get an estimate for the typical predictions of
the schemes with heavy Majorana neutrinos 
with hierarchical spectrum we will consider further, 
we introduce a ``benchmark SUSY scenario'' defined by the 
values of the SUSY parameters 
%%%%%%%%%%%%%%%%%%%%%%%%%%%%%%%%%%%%%%%%%%
\begin{align}
m_0 = m_{1/2} = 250~{\rm GeV},~~~A_0 = a_0m_0 = - 100~{\rm GeV}\;,
\label{bench}
\end{align}
%%%%%%%%%%%%%%%%%%%%%%%%%%%%%%%%%%%%%%%%%%
%
and $\tan\beta \sim (5 - 10)$.  In this scenario the lightest
supersymmetric particle is a neutralino with a mass of $\sim 100$ GeV.
The next to the lightest SUSY particles are the chargino and a second
neutralino with masses $\sim 200$ GeV. The squarks have masses in the
range of $\sim (400 - 600)$ GeV.  Supersymmetric particles possessing
the indicated masses can be observed in the experiments under
preparation at the LHC.

   The ``benchmark'' values of $m_0$, $m_{1/2}$ and  $A_0$  
in eq.~(\ref{bench}) correspond to
%%%%%%%%%%%%%%%%%%%%%%%%%%%%%%%%%%%%%%%%%%
\begin{equation} 
\label{eq:bench}
\text{B}(l_i \to l_j + \gamma) \simeq 9.1 \times 10^{-10} 
\left| \left(\mathbf{Y_{\nu}^\dagger} L \mathbf{Y_{\nu}}\right)_{ij} \right|^2 
\, \tan^2 \beta~,
\end{equation}
%%%%%%%%%%%%%%%%%%%%%%%%%%%%%%%%%%%%%%%%%%%
%
where $(L)_{kl} = \delta_{kl}(L)_{k}$, $L_k \equiv \ln(M_{X}/M_k)$.
Since $\tan^2 \beta$ will typically enhance $\text{B}(l_i \to l_j +
\gamma)$ by at least 1 order of magnitude, the quantity
$\left|\left(\mathbf{Y_{\nu}^\dagger} L
    \mathbf{Y_{\nu}}\right)_{21}\right|$ has to be relatively small to
be in agreement with the existing experimental upper limit on
$\text{B}(\mu \to e + \gamma)$. For given values of the heavy Majorana
neutrino masses, this will lead to certain constraints on the
parameters in $\mathbf{R}$. Regarding the masses of the heavy Majorana
neutrinos, we shall assume that $M_1 \ll M_2 \ll M_3$, with $M_3$
having a value $M_3 \gtap (10^{13} - 10^{14})$ GeV, $M_3 \ll M_X$.
Constraints from thermal leptogenesis require that $M_1 \gs 10^9$
GeV~\cite{LGBDiBP05,CERN04}.  This would indicate a hierarchy, $\eg$,
of the form $M_1 \simeq (10^{9}-10^{11})$ GeV,
$M_2\simeq(10^{12}-10^{13})$ GeV and $M_3\simeq(10^{14}-10^{15})$ GeV.
 
%%%%%%%%%%%%%%%%%%%%%%%%%%%%%%%%%%%%%%%%%
%
\subsection{\label{sec:YB}\large{Leptogenesis}} 
%
%%%%%%%%%%%%%%%%%%%%%%%%%%%%%%%%%%%%%%%%%%

\indent\ In the case of $M_1 \ll M_2 \ll M_3$,
the baryon asymmetry of interest is given by 
%%%%%%%%%%%%%%%%%%%%%%%%%%%%%%%%%%%%%%%%%%%
\begin{equation}
Y_B  \simeq - 10^{-2}\, \kappa\, \epsilon_1\;,
\label{YBobsth}
\end{equation}
%%%%%%%%%%%%%%%%%%%%%%%%%%%%%%%%%%%%%%%%%%%
%
where $\epsilon_1$ is the CP-violating asymmetry in the decay of the
lightest RH Majorana neutrino $N_1$ having the mass $M_1$, and
$\kappa$ is an efficiency factor calculated by solving the Boltzmann
equations (see, $\eg$,~\cite{LGBDiBP05}).  A simple approximate
expression for the efficiency factor $\kappa$ in the case of thermal
leptogenesis we will assume in what follows, was found in~\cite{CERN04}:
%%%%%%%%%%%%%%%%%%%%%%%%%%
\begin{align}
\frac{1}{\kappa} \simeq \frac{3.3\times 10^{-3}\text{eV}}{\widetilde{m}_1}
+\left(\frac{\widetilde{m}_1}{0.55\times 10^{-3}\text{eV}}\right)^{1.16}\;,
\end{align}
%%%%%%%%%%%%%%%%%%%%%%%%%%%
%
where the neutrino mass parameter $\widetilde{m}_1$ is given by 
%%%%%%%%%%%%%%%%%%%%%%%%%%%%%%%%%%%%
\begin{align}
\widetilde{m}_1 \equiv \frac{v_u^2}{M_1}\, \left(\mathbf{Y_{\nu}} \mathbf{Y_{\nu}^{\dagger}}\right)_{11}\;.
\label{tilm1}
\end{align}
%%%%%%%%%%%%%%%%%%%%%%%%%%%
%
The CP-violating decay asymmetry $\epsilon_1$ has the form
%%%%%%%%%%%%%%%%%%%%%%%%%%%%%%%%%%%%%%%%%%
\begin{equation}
\epsilon_1 
\simeq - \frac{3}{8\, \pi} 
\, \frac{1}{\left(\mathbf{Y_{\nu}} \, \mathbf{Y_{\nu}^\dagger}\right)_{11}} 
\, {\rm Im} \left\{ \left(\mathbf{Y_{\nu}} \, 
  \mathbf{Y_{\nu}^\dagger}\right)_{21}^2 \right\} \, 
\frac{M_1}{M_2}~. 
\label{e1H}
\end{equation}
%%%%%%%%%%%%%%%%%%%%%%%%%%%%%%%%%%%%%%%%%%%
%
Extensive numerical studies have shown~\cite{LGBDiBP05,CERN04} that in
MSSM and for hierarchical spectrum of masses of the heavy Majorana
neutrinos under discussion, successful thermal leptogenesis is
possible only for
%%%%%%%%%%%%%%%%%%%%%%%%%%
\begin{align}
\widetilde{m}_1 \ltap 0.12~{\rm eV}\;.
\label{maxtilm1}
\end{align}
%%%%%%%%%%%%%%%%%%%%%%%%%%%
%
{}For typical values of $\kappa \sim (10^{-3} - 10^{-1})$ one gets for $Y_B$ 
a value compatible  with the observations~\cite{WMAPnu},
%%%%%%%%%%%%%%%%%%%%%%%%%%%%%
\begin{align}
Y_B = (6.15 \pm 0.25)\times 10^{-10}\;,
\label{YBobs}
\end{align}
%%%%%%%%%%%%%%%%%%%%%%%%%%%%
%
if $\epsilon_1\sim -(10^{-5}-10^{-7})$. 

\indent\ As it follows from eqs.~(\ref{YnudYnu}) and~(\ref{eq_ijg}),
the branching ratios of $l_i \to l_j + \gamma$ decays in the case of
interest depend on the orthogonal matrix $\mathbf{R}$. Successful
leptogenesis can take place only if $\mathbf{R}$ is complex, so we
will consider $(\mathbf{R})^{*}\neq \mathbf{R}$.  In what follows we
will use a parametrisation of $\mathbf{R}$ with complex angles (see,
$\eg$,~\cite{Iba01,JohnE}):
%%%%%%%%%%%%%%%%%%%%%%%%%%%%%%%%%%%%%
\begin{align}
\mathbf{R} = \mathbf{R}_{12} \, \mathbf{R}_{13} \, \mathbf{R}_{23}~,
~~{\rm or}~~
\mathbf{R} = \mathbf{R}_{12} \, \mathbf{R}_{23} \, \mathbf{R'}_{12}~, 
\label{R3rot}
\end{align}
%%%%%%%%%%%%%%%%%%%%%%%%%%%%%%%%%%%%%
%
where $\mathbf{R}_{ij}$ ($\mathbf{R'}_{12}$) describes now the
rotation with a complex angle $\omega_{ij} = \rho_{ij} + i\sigma_{ij}$
($\omega'_{12} = \rho'_{12} + i\sigma'_{12}$), $\rho_{ij}$ and
$\sigma_{ij}$ ($\rho'_{12}$ and $\sigma'_{12}$) being real parameters.
These parametrisations prove particularly convenient for investigating
the case of hierarchical spectrum of masses of the heavy RH neutrinos.

%%%%%%%%%%%%%%%%%%%%%%%%%%%%%%%%%%%%%%%%%%%%%%%%%%%
%
\section{\large{The See-Saw Mechanism, Neutrino Yukawa Couplings, 
LFV Decays $l_i \to l_j +\gamma$ and Leptogenesis}}
%
%%%%%%%%%%%%%%%%%%%%%%%%%%%%%%%%%%%%%%%%%%%%%%%%%%%
%
\indent\ There has been a considerable theoretical effort in recent
years to understand possible connections between the neutrino mass and
mixing data, LFV charged lepton decays and leptogenesis.  Here we
shall focus on the combined constraints which the existing stringent
upper limit on the $\mu \to e + \gamma$ decay rate and the requirement
of successful leptogenesis impose on MSSMRN. We will be interested, in
particular, in the possible implications of these constraints for the
form of the matrix $\mathbf{R}$, the heavy Majorana neutrino masses,
the predicted rates of the decays $\mu \to e + \gamma$, $\tau \to e +
\gamma$ and $\tau \to \mu + \gamma$, and the basic SUSY parameters.

%%%%%%%%%%%%%%%%%%%%%%%%%%%%%%%%%%%%%%%%%%%
%
\subsection{\label{sec:NHNH}
\large{Normal Hierarchical Light Neutrino Mass Spectrum}}
%
%%%%%%%%%%%%%%%%%%%%%%%%%%%%%%%%%%%%%%%%%%

\indent\ We set first $m_1$, $M_1$ and $M_2$ to zero.  In this
approximation we find using $\mathbf{R}=
\mathbf{R}_{12}\mathbf{R}_{13}\mathbf{R}_{23}$:
%%%%%%%%%%%%%%%%%%%%%%%%%%%%%%%%%%%%%%%%%
\begin{eqnarray}
\left(\mathbf{Y_{\nu}^\dagger} L \mathbf{Y_{\nu}}\right)_{21} &\simeq&
\frac{\D L_3\,M_3 \, \cos \omega_{13}\,
   \cos \omega_{13}^*}{\D \sqrt{2} \, v_u^2} \nonumber \\ %[0.3cm]
&& \times \,\bigg[ \sqrt{m_2}\, \left( e^{-i \,\left( \alpha - \beta_M
     \right)/2 }\, \sqrt{m_3}\, \cos \omega_{23}^* - \sqrt{m_2}
   \,c_{12}\,
   \sin \omega_{23}^* \right)\,s_{12}\,\sin \omega_{23} \nonumber\\%[0.3cm]
&&  - m_3 \,\cos \omega_{23}\,\cos \omega_{23}^*\, s_{13}\,e^{i \delta}
\bigg]~.
\label{21NHM3}
\end{eqnarray}
%%%%%%%%%%%%%%%%%%%%%%%%%%%%%%%%%%%%%%%%%%%%%%%
%
In deriving eq.~(\ref{21NHM3}) we have set for simplicity $\theta_{23}
= \pi/4$ and neglected the terms $\propto m_2s_{13}$ and $\propto
m_{2,3} s^2_{13}$ in the square brackets.  The corresponding
expressions for $\left(\mathbf{Y_{\nu}^\dagger} L
  \mathbf{Y_{\nu}}\right)_{31, 32}$ are very similar to that for
$\left(\mathbf{Y_{\nu}^\dagger} L \mathbf{Y_{\nu}}\right)_{21}$.  In
particular, both are proportional to $ L_3\,M_3 \, \cos \omega_{13}\,
\cos \omega_{13}^\ast$.  For the plausible values of $M_3 \cong
(10^{14}-10^{15})$ GeV and $M_X \cong 2\times 10^{16}$ GeV, one finds
that $M_3\,L_3\,\sqrt{\dma}/(\sqrt{2}v_u^2) \gtap 0.66$.  Barring
accidental cancellations between the terms in the square brackets 
in eq.~(\ref{21NHM3}), we get from eq.~(\ref{eq:bench}) that the
predicted value of $\text{B}(\mu \to e + \gamma)$ will be larger at
least by a factor of $\sim 10^3$ than the existing upper bound
$\text{B}(\mu \to e + \gamma)$.  For $M_3 \cong 10^{13}$ GeV and $M_X =
2\times 10^{16}$ GeV one finds $M_3\,L_3\,\sqrt{\dma}/(\sqrt{2}v_u^2)
\cong 0.09$, which for $\tan^2\beta = 25~(100)$ and the chosen
``benchmark'' values of the SUSY parameters leads to $\text{B}(\mu \to e +
\gamma) \cong 1.8 \times 10^{-10}~(7.4 \times 10^{-10}$). 
The values obtained are 
still larger than the current limit.  
This might suggest that the SUSY
parameter $m_{1/2}$ has a bigger value than the ``benchmark scenario''
value we have assumed, or that $m_0 \sim m_{1/2} \gtap 500$
GeV~\footnote{Information about the SUSY parameters $m_0$ and
  $m_{1/2}$ of interest is expected to be obtained in the experiments
  under preparation at the LHC.}.  We will pursue, however, an
alternative hypothesis.  We will suppose that $m_{1/2} \sim {\rm
  few}\times 100$ GeV.  If indeed $M_3 \gtap 5\times 10^{13}$ GeV,
$M_3 \ll M_{\rm X}$, where $M_{\rm X} \geq M_{\rm GUT}$, the existing
stringent experimental upper limit on $\text{B}(\mu \to e + \gamma)$
might suggest that $\omega_{13} \cong \pi/2$ and we will explore this
interesting possibility in what follows.  For $\omega_{13} = \pi/2$,
the $\mathbf{R}$ matrix has the form
%%%%%%%%%%%%%%%%%%%%%%%%%%%%%%%%%%%%%%%%
\bea
\mathbf{R} \simeq 
\left( 
\bad 
0 & \sin\omega & \cos\omega \\
0 & \cos\omega & -\sin\omega \\
-1 & 0 & 0 
\ea 
\right)~~, 
\label{RNH}
\eea
%%%%%%%%%%%%%%%%%%%%%%%%%%%%%%%%%%%%%%%%%%
%
where $\omega \equiv \omega_{12} - \omega_{23}$.  Thus, only the
combination $(\omega_{12} - \omega_{23})$ of the complex angles
$\omega_{12}$ and $\omega_{23}$ appears in the expression for
$\mathbf{R}$.  It is not difficult to convince oneself that if $m_1$
is negligible, and $\mathbf{R}$ has the form given in eq.~(\ref{RNH}),
we have $(\mathbf{Y_{\nu}})_{3j} = 0~(j=1,2,3)$.  This means that the
heaviest RH Majorana neutrino $N_3$ decouples and $\mathbf{Y}_{\nu}$
coincides in form with the matrix of neutrino Yukawa couplings in the
so-called ``3$\times$2'' see-saw model~\cite{FGY03}. 
%%%%%%%%%%%%%%%%%%%%%%%%%%%%%%%%%%%%%%%%%%
%% comment outed by T.S.
We will keep,
however, the elements $(\mathbf{Y_{\nu}})_{1j(2j)} \neq 0~(j=1,2,3)$ in 
our further analysis.
%%%%%%%%%%%%%%%%%%%%%%%%%%%%%%%%%%%%%%%%%%

With $m_1 = 0$ and $\mathbf{R}$ having the form~(\ref{RNH}), the terms
$\sim M_2L_2$ give the dominant contribution in
$\left|\left(\mathbf{Y_{\nu}^\dagger} L \mathbf{Y_{\nu}}\right)_{ij}\right|~(i\neq j)$.  We
get:
%%%%%%%%%%%%%%%%%%%%%%%%%%%%%%%%%%%%%%%%%%%%%
\begin{eqnarray}
\label{eq:nhnh21} 
\left| \left(\mathbf{Y_{\nu}^\dagger} L \mathbf{Y_{\nu}}\right)_{21} \right|
&\simeq& \frac{\D L_2\,M_2}{\D v_u^2} \left| \left( e^{-i \alpha/2} \,
     \sqrt{m_2} \, c_{13} \, s_{12} \, c_\omega - e^{-i (\beta_M - 2
       \delta)/2} \, \sqrt{m_3} \, s_{13} \, s_\omega
   \right) \right. \nonumber\\ 
&& \times \left. \left( e^{i \beta_M/2} \, \sqrt{m_3} \,c_{13}\, s_{23} \,
     s_\omega^* - e^{i \alpha/2} \, \sqrt{m_2} \, c_{12} \, c_{23}\,
     c_\omega^* \right) \right| ~, 
\end{eqnarray}
%%%%%%%%%%%%%%%%%%%%%%%%%%%%%%%%%%%%%%%%%%%%%
%
where $c_\omega \equiv \cos \omega$, $s_\omega \equiv \sin \omega$ and
we have neglected terms $\propto s_{13}$ which can give a correction
not exceeding approximately $13\%$. Setting, for instance $M_2 =
10^{12}$ GeV, we find $|(\mathbf{Y_{\nu}^\dagger} L
\mathbf{Y_{\nu}})_{21}| \sim M_2 \, L_2\, \sqrt{\dma}/v_u^2 \simeq
10^{-2}$, and correspondingly $\text{B}(\mu \to e + \gamma) \cong
2.3\times 10^{-12}$ for $\tan^2\beta = 25$. This is the range that
will be explored by the experiment MEG~\cite{psi} currently under
preparation.  Similarly, we obtain for $\left(\mathbf{Y_{\nu}^\dagger} L
\mathbf{Y_{\nu}}\right)_{31, 32}$:
%%%%%%%%%%%%%%%%%%%%%%%%%%%%%%%%%%%%%%%%%%%%%%%
\begin{eqnarray}
\label{eq:nhnh31}
\left| \left(\mathbf{Y_{\nu}^\dagger} L \mathbf{Y_{\nu}}\right)_{31} \right|
&\simeq& \frac{\D L_2\,M_2}{\D v_u^2} 
\left|\left( e^{-i \alpha/2} \, \sqrt{m_2} \, s_{12} \, c_\omega - 
e^{-i (\beta_M - 2 \delta)/2} \, \sqrt{m_3} \, s_{13} \, s_\omega 
\right) \right. \nonumber\\ 
&& \times\left. \left( e^{i \beta_M/2} \, \sqrt{m_3} \, c_{23} \, s_\omega^* + 
e^{i \alpha/2} \, \sqrt{m_2} \, 
c_{12} \, s_{23} \, c_\omega^* \right)\right|\;,
\end{eqnarray}
%%%%%%%%%%%%%%%%%%%%%%%%%%%%%%%%%%%%%%%%%%
%
and
%%%%%%%%%%%%%%%%%%%%%%%%%%%%%%%%%%%%%%%%%%
\begin{eqnarray}
\label{eq:nhnh32}
\left| \left(\mathbf{Y_{\nu}^\dagger} L \mathbf{Y_{\nu}}
\right)_{32}\right|
  &\simeq& \frac{\D L_2\,M_2}{\D v_u^2} 
\left| \left( 
e^{-i \alpha/2} \, \sqrt{m_2} \, 
c_{12} \, c_{23}\, c_\omega + 
e^{-i \beta_M/2} \, \sqrt{m_3} \, s_{23} \, s_\omega 
\right)\right. \nonumber\\ 
&& \times \left. \left( 
e^{i \beta_M/2} \, \sqrt{m_3} \, c_{23} \, s_\omega^* 
+ e^{i \alpha/2} \, \sqrt{m_2} \, c_{12} \, s_{23} 
\, c_\omega^* 
\right)\right|\;. 
\end{eqnarray}
%%%%%%%%%%%%%%%%%%%%%%%%%%%%%%%%%%%%%%%%%%%%%%
%
As it follows from eqs.~(\ref{eq:nhnh21})--(\ref{eq:nhnh32}), for
$\omega \neq 0$, the $l_i \to l_j +\gamma$ decay branching ratios of
interest depend on the Majorana phase difference $(\alpha-\beta_M)$.
The effective Majorana mass $\meff$ in $\betabeta$-decay 
(see, $\eg$,~\cite{BiPet87,APSbb0nu}) depends on the same 
Majorana phase difference (see, $\eg$,~\cite{BPP1,STPFocusNu04}):
%%%%%%%%%%%%%%%%%%%%%%%%%%%%%%%%%%
\begin{equation}
\meff \cong \left|\sqrt{\deltasol}\sin^2 \theta_{12} e^{i(\alpha-\beta_{M})}
+ \sqrt{\deltaatm}\sin^2 \theta_{13} \right|\;.
\label{meffNH2}
\end{equation}
%%%%%%%%%%%%%%%%%%%%%%%%%%%%%%%
%
If $s_{13}|s_{\omega}|$ is not negligibly small, $\text{B}(\mu \to e +
\gamma)$ and $\text{B}(\tau \to e + \gamma)$ will depend also on the
Dirac phase $\delta$.

For the ``double'' ratio $\text{R}(21/31)$ 
we find from  eqs.~(\ref{eq:nhnh21}) and~(\ref{eq:nhnh31}):
%%%%%%%%%%%%%%%%%%%%%%%%%%%%%%
\begin{align}
\text{R}(21/31) \cong 
\frac{\left |\sqrt{m_3} \, s_{23} \, s_\omega 
- \sqrt{m_2}\, c_{12} \, c_{23}\, c_\omega \,
e^{i \frac{\alpha - \beta_M}{2}}\,\right |^2} 
{\left |\sqrt{m_3} \, c_{23} \, s_\omega 
+ \sqrt{m_2}\, c_{12} \, s_{23}\, c_\omega 
e^{i \frac{\alpha - \beta_M}{2}}\,\right |^2}\;.
\label{R2132NHNHM2}
\end{align}
%%%%%%%%%%%%%%%%%%%%%%%%%%%%%%
%
Given $\theta_{12}$, $\theta_{23}$ and $m_2/m_3 \cong
\sqrt{\deltasol}/\sqrt{\deltaatm}$, $\text{R}(21/31)$ depends only on
$(\alpha-\beta_M)$ and $\omega$. If the terms $\propto \sqrt{m_3}$
($\propto \sqrt{m_2}$) in eq.~(\ref{R2132NHNHM2}) dominate, we have
$\text{R}(21/31) \cong 1$.
 
The expression for the double ratio $\text{R}(21/32)$ can be
obtained from eqs.~(\ref{eq:nhnh21}) and~(\ref{eq:nhnh32}).  For
$\sqrt{m_3} s_{13} |s_\omega| \gg \sqrt{m_2}|c_\omega|$ we get
$\text{R}(21/32) \cong s^2_{13}/c^2_{23} \ltap 0.1$, while if
$\sqrt{m_3} |s_\omega| \ll \sqrt{m_2}|c_\omega|s_{12}$, one finds
$\text{R}(21/32) \cong (\tan^2\theta_{12})/s^{2}_{23} \cong 0.9$.

%%%%%%%%%%%%%%%%%%%%%%%%%%%%%%%%%%%%%%%%%
%
\subsubsection{\label{sec:NHNH_YB}\large{Leptogenesis Constraints}} 
%
%%%%%%%%%%%%%%%%%%%%%%%%%%%%%%%%%%%%%%%%%%

\indent We shall analyse next the constraints on the parameter
$\omega$ in the matrix $\mathbf{R}$, eq.~(\ref{RNH}), which follow
from the requirement of successful thermal leptogenesis.  With
$\omega_{13} = \pi/2$ we find that in the case of NH light neutrino
mass spectrum we are considering.
%%%%%%%%%%%%%%%%%%%%%%%%%%%%%%%%%%%%%%%%
\begin{align} 
\label{eq:epsNH1}
\epsilon_1 & \simeq 
- \frac{3}{8\pi} \left (\frac{m_3M_1}{v_u^2}\right )
\frac{{\rm Im}\left [c^2_{\omega} + 
\frac{\deltasol}{\deltaatm}~s^2_{\omega}\right ]}
{|c_{\omega}|^2 + 
\sqrt{\frac{\deltasol}{\deltaatm}}~|s_{\omega}|^2} \\[0.3cm]
\label{eq:epsNH2}
& \simeq
\frac{3}{8\pi} \left (\frac{m_3M_1}{v_u^2}\right )
\frac{\sin 2 \rho \, \sinh 2 \sigma} 
{(1 - \frac{m_2}{m_3})\cos 2 \rho 
+ (1 + \frac{m_2}{m_3}) \cosh 2 \sigma }~,
\end{align}
%%%%%%%%%%%%%%%%%%%%%%%%%%%%%%%%
%
where $\rho$ and $\sigma$ are determined by $\omega = \rho + i
\sigma$ and we have used $m_2^2\cong \deltasol$, $m_3^2\cong
\deltaatm$ and the relation ${\rm Im}c^2_{\omega} = -{\rm
  Im}s^2_{\omega}$.  Thus, in the case under discussion, the mass
$M_2$ governs the magnitude of the $l_i \to l_j + \gamma$ decay
branching ratios, whereas $M_1$ determines the value of the
leptogenesis decay asymmetry.  The conditions $|c_{\omega}|^2 \geq 0$,
$|s_{\omega}|^2\geq 0$ imply that $\rho$ and $\sigma$ 
(in eq.~(\ref{eq:epsNH2})) should satisfy $\cosh 2\sigma \geq |\cos 2 \rho|$,
which is always valid since $\cosh 2\sigma \geq 1$.  As can be easily
shown, we have
%%%%%%%%%%%%%%%%%%%%%%%%%%%%%%%%%%%%%%%%%%
\begin{equation}
 \frac{\left | {\rm Im}\left[ c^2_{\omega} + 
\frac{\deltasol}{\deltaatm}~s^2_{\omega} \right]\right|}
{\left|c_{\omega}\right|^2 + 
\sqrt{\frac{\deltasol}{\deltaatm}}~\left|s_{\omega}\right|^2} 
\leq  \frac{\left| {\rm Im}c^2_{\omega}\right|}
{\left|c_{\omega}\right|^2}  \leq 1\;,
\end{equation}
%%%%%%%%%%%%%%%%%%%%%%%%%%%%%%%%%%%%%%%%%%%
%
leading to the well-known~\cite{daviba02} upper limit
%%%%%%%%%%%%%%%%%%%%%%%%%%%%%%%%%%%%%%%%%%%
\begin{equation}
\left|\epsilon_1\right| \ltap \frac{3}{8\pi} \left (\frac{m_3M_1}{v_u^2}\right )
\simeq 1.97\times 10^{-7}
\left (\frac{m_3}{0.05~{\rm eV}}\right )
\left (\frac{M_1}{10^{9}~{\rm GeV}}\right )
\left (\frac{174~{\rm GeV}}{v_u}\right )^2\;.
\label{maxasymNH}
\end{equation}
%%%%%%%%%%%%%%%%%%%%%%%%%%%%%%%%%%%%%%%%%%%
%
The requirement of a nonzero asymmetry, $\epsilon_1 \neq 0$,
implies, as it follows from eq.~(\ref{eq:epsNH2}), that both the real
and imaginary parts of $\omega$ have to be nonzero, $\rho\neq k\pi/2$,
$k=0,1,2,...$, $\sigma\neq 0$, $\ie$, that $\mathbf{R}$ has to be
complex.  Moreover, we should have $\sin 2\rho\sinh 2\sigma < 0$ since
the decay asymmetry $\epsilon_1$ has to be negative in order to
generate a baryon asymmetry of the correct sign.  The maximal
asymmetry $\left|\epsilon_1\right|$ is obtained for $|{\rm Im}c^2_{\omega}| =
|c_{\omega}|^2$, which is satisfied for $\cos 2\rho~\cosh 2\sigma =
-1$, $\cos 2\rho \neq -1$, $\cosh 2\sigma \neq 1$.

The neutrino mass parameter $\widetilde{m}_1$, eq.~(\ref{tilm1}), can also
be easily found:
%%%%%%%%%%%%%%%%%%%%%%%%%%%%%%%%%%%
\begin{align}
\widetilde{m}_1 \simeq m_3|c_{\omega}|^2 + m_2|s_{\omega}|^2
= \frac{1}{2}(m_3+m_2)\cosh2\sigma
 + \frac{1}{2}(m_3-m_2)\cos 2\rho \geq m_2\;. 
%\sim 9\times 10^{-3}~{\rm eV}\;.
\label{tilm1NH}
\end{align}
%%%%%%%%%%%%%%%%%%%%%%%%%%%%%%%%%%%%%%%%%%%%%%%%%%%%%%%
%%%%
%%% Figure 1
%%%%
\begin{figure}[!t]
\begin{center}
%%%%%
\begin{tabular}{cc}
\includegraphics{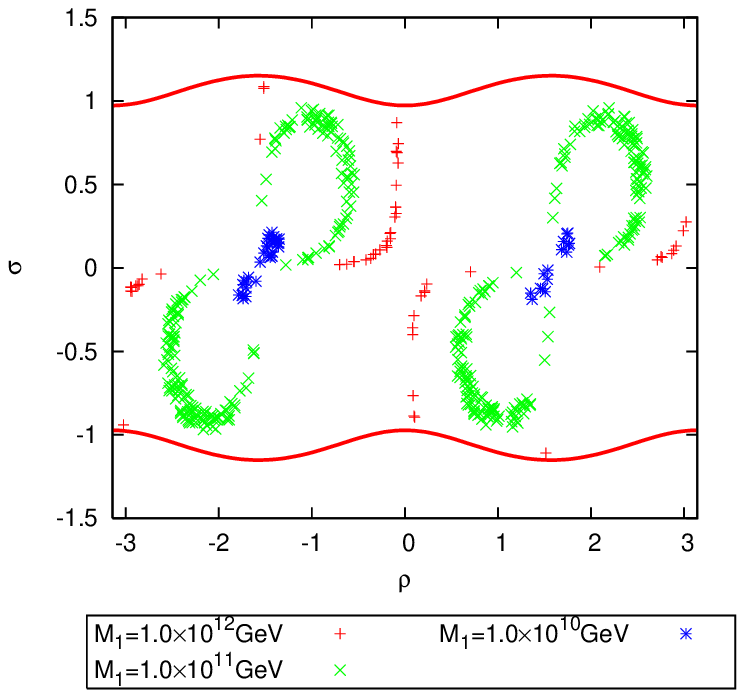}&
\includegraphics{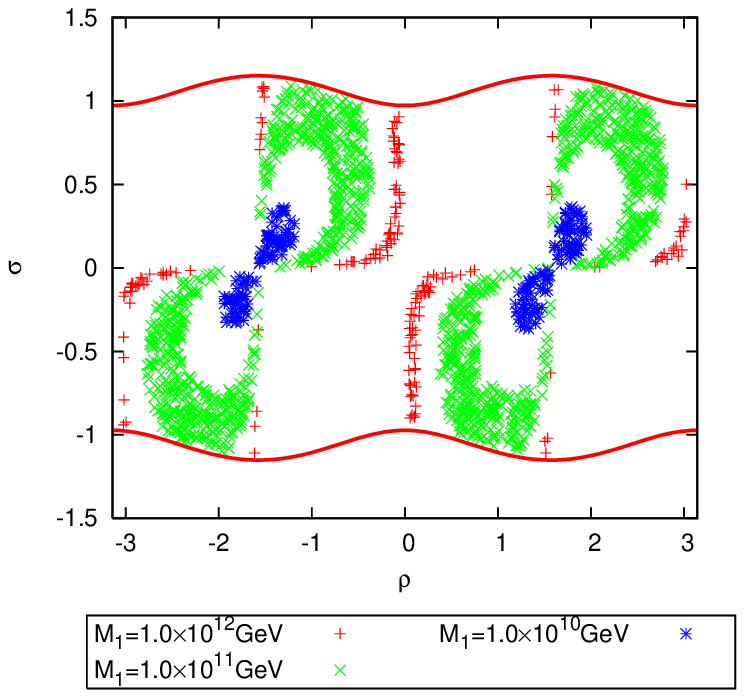}\\
(a)&(b)
\end{tabular}
%%%%%
\caption{The leptogenesis constraints on  
  the parameters $\rho$ and $\sigma$ for $M_1 = 10^{10}$ GeV;
  $10^{11}$ GeV; $10^{12}$ GeV (blue; green; red areas) in the case of
  NH light neutrino mass spectrum.  The two panels correspond to two
  different intervals of values of the baryon asymmetry of the
  Universe, $Y_B$, considered: (a) $5.0\times 10^{-10}\leq Y_B\leq
  7.0\times 10^{-10}$, and (b) $3.0\times 10^{-10}\leq Y_B\leq
  9.0\times 10^{-10}$.  The solid lines show the limit associated with
  the upper bound $\widetilde{m}_1\leq 0.12$ eV: outside the region
  between the solid lines the wash-out effects are too strong and
  leptogenesis cannot produce the observed baryon asymmetry.}
\label{rho-sigma-nh}
\end{center}
\end{figure}
%
%%%%%%%%%%%%%%%%%%%%%%%%%%%%%%%%%%%%%%%%%%%%%%%%%%%%%%%
The minimal value of $\widetilde{m}_1 = m_2 \cong \sqrt{\deltasol} \cong
9\times 10^{-3}~{\rm eV}$, corresponds to $\cosh2\sigma = 1$ and $\cos
2\rho = - 1$, for which $\left|\epsilon_1\right| = 0$.  For $9\times 10^{-3}~{\rm
  eV} < \widetilde{m}_1 \ltap 0.12$ eV, where we have taken into account 
eq.~(\ref{maxtilm1}), the efficiency factor lies in the interval
$1.9\times 10^{-3} \ltap \kappa < 3.9\times 10^{-2}$.  For this range
of values of $\kappa$ successful leptogenesis is possible for $M_1
\gtap 10^{10}$ GeV. We will consider values of $M_1$ in the interval
$M_1 = (10^{10} - 10^{12})$ GeV, which is compatible with the
assumption we made about the hierarchical mass spectrum of the heavy
Majorana neutrinos.  Thus, for given $M_1$, the requirement of
successful leptogenesis implies a constraint on the two parameters
$\rho$ and $\sigma$ of the theory.  In Fig.~\ref{rho-sigma-nh} we show
the leptogenesis constraint on $\rho$ and $\sigma$ for 
$M_1= 10^{10}$ GeV; $10^{11}$ GeV; $10^{12}$ GeV.  As we see from
Fig.~\ref{rho-sigma-nh}, the requirement of successful leptogenesis
severely limits the allowed ranges of values of $\rho$ and $\sigma$.
Moreover, the values of the two parameters are strongly correlated. We
note, in particular, that as $|\sigma|$ increases, the wash-out effects
become stronger and for $|\sigma| \gtap 1$, the observed baryon
asymmetry cannot be reproduced.  The maximal asymmetry $\left|\epsilon_1\right|$
for a given $M_1$ is obtained for values of $\cos 2\rho$ and $\cosh
2\sigma$ close, but not equal, to $(-1)$ and $(+1)$, respectively.
%
%%%%%%%%%%%%%%%%%%%%%%%%%%%%%%%%%%%%%%%%%%%%%%%%%%%%%%%
%%%%
%%% Figure 2
%%%%
\begin{figure}[!t]
\begin{center}
%%%%%
\begin{tabular}{cc}
\includegraphics{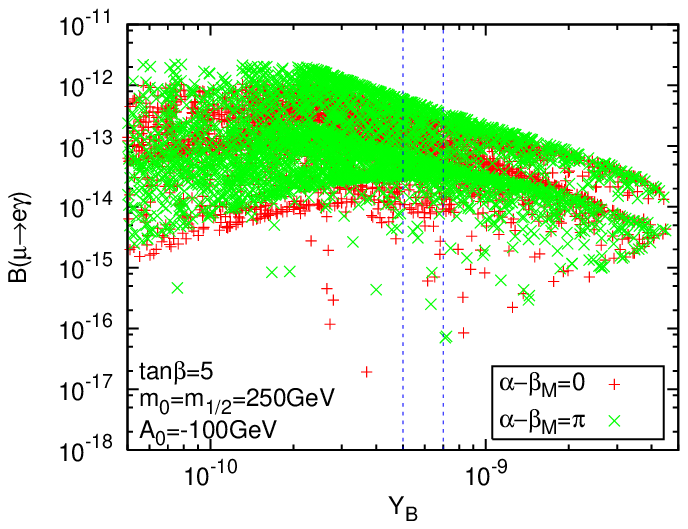}&
\includegraphics{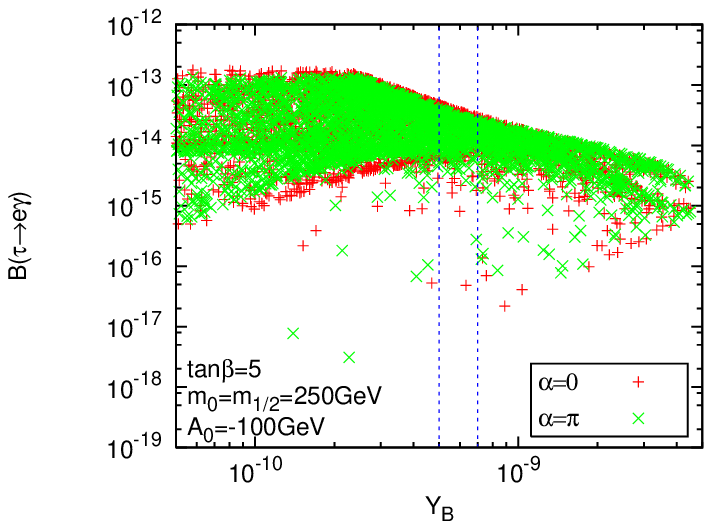}\\
(a)&(b)\\
\includegraphics{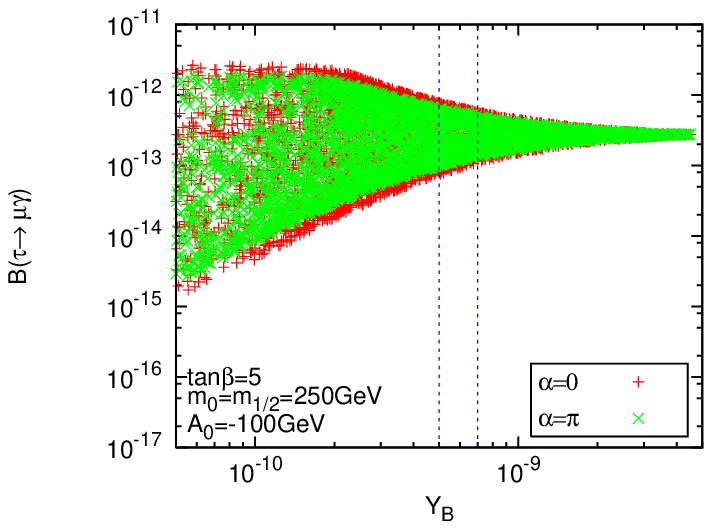}&\\
(c)&
\end{tabular}
%%%%%
\caption{The correlation between 
  the predicted values of the LFV decay branching ratios
  $\text{B}(\mu\to e+\gamma)$ (a), $\text{B}(\tau\to e+\gamma)$ (b),
  $\text{B}(\tau\to \mu+\gamma)$ (c) and of the baryon asymmetry $Y_B$
  in the thermal leptogenesis scenario, for $M_1 = 6\times 10^{10}$
  GeV, $M_2 = 10^{12}$ GeV and NH light neutrino mass spectrum.  The
  figure was obtained for the ``benchmark'' values of the soft SUSY
  breaking parameters $m_0 = m_{1/2} = 250$ GeV, $a_0m_0 = -100$ GeV
  and the minimal value of $\tan\beta = 5$.  The region between the
  two vertical dashed lines corresponds to the observed baryon
  asymmetry: $5.0\times 10^{-10}\leq Y_B\leq 7.0\times 10^{-10}$.
  Results for two values of the Majorana phase $(\alpha - \beta_M)$
  equal to 0 (red+green areas) and $\pi$ (green areas) are shown.  }
\label{etaB-lfv-nh}
\end{center}
\end{figure}
%%%%%%%%%%%%%%%%%%%%%%%%%%%%%%%%%%%%%%%%%%%%%%%%%%%%%%%
%
Given the interval of allowed values of $\widetilde{m}_1$, we can write
$\widetilde{m}_1 = f\,m_3$ with $f \cong [0.2,2]$. The condition of
maximal $\left|\epsilon_1\right|$ implies $\cos 2\rho = (f - \sqrt{1 +
  f^2})m_3/(m_3 - m_2)$.  Choosing $f = 1$, $\ie$, $\widetilde{m}_1 = m_3
\cong 0.05$ eV, for instance, we get $\cos 2\rho \cong -0.5$ and
correspondingly $\cosh 2\sigma \cong 2$. In this example $\kappa \cong
5.4\times 10^{-3}$ and for $\tan^2\beta \gtap 3$ we get the requisite
value of the the baryon asymmetry for $M_1 \cong 6\times 10^{10}$ GeV.

In Fig.~\ref{etaB-lfv-nh} we show the relation between the predicted
values of $Y_B$ in the thermal leptogenesis scenario and of
$\text{B}(l_i\to l_j + \gamma)$.  The figure was obtained for the
``benchmark'' values of the soft SUSY breaking parameters $m_0 =
m_{1/2} = 250$ GeV, $a_0m_0 = -100$ GeV and the minimal value of
$\tan\beta = 5$. For the relevant heavy Majorana neutrino masses we
used $M_1 = 6\times 10^{10}$ GeV and $M_2 = 10^{12}$ GeV. Results for
two values of the Majorana phase difference $(\alpha-\beta_M) =
0;~\pi$, are shown. For the values of $\rho$ and $\sigma$ ensuring
successful leptogenesis we find that typically $10^{-14} \ltap
\text{B}(\mu\to e +\gamma)\ltap 5\times 10^{-13}$, $10^{-13} \ltap
\text{B}(\tau\to \mu + \gamma)\ltap 5\times 10^{-12}$ and $10^{-15}
\ltap \text{B}(\tau\to e + \gamma)\ltap 5\times 10^{-13}$.  However,
we have $\text{B}(l_i\to l_j +\gamma)\propto\tan^2\beta$ and, $\eg$,
for $\tan\beta = 20$ we get typically $1.6\times 10^{-13} \ltap
\text{B}(\mu\to e +\gamma)\ltap 8\times 10^{-12}$, which is entirely
in the range of sensitivity of the MEG experiment. As Fig.~\ref{etaB-lfv-nh} 
indicates, the dependence of $\text{B}(l_i\to l_j +\gamma)$ on the 
Majorana phase $(\alpha-\beta_M)$ is relatively weak.

%%%%%%%%%%%%%%%%%%%%%%%%%%%%%%%%%%%%%%%%%%%
%
\subsection{\label{sec:NHIH}
\large{Inverted Hierarchical Light Neutrino Mass Spectrum}}
%
%%%%%%%%%%%%%%%%%%%%%%%%%%%%%%%%%%%%%%%%%%

\indent\ We will perform next a similar analysis assuming that the
light neutrino mass spectrum is of the inverted hierarchical type.  We
set $m_2 \cong m_1$, $m_3/m_{1,2} \cong 0$ and neglect first $M_1/M_3$
and $M_2/M_3$.  Setting for simplicity $s_{13}=0$ and $\theta_{23} =
\pi/4$, we find that $\text{B}(\mu \to e + \gamma)$ will depend on
%%%%%%%%%%%%%%%%%%%%%%%%%%%%%%%%%%%%%%%%%%%%%
\begin{eqnarray}
  \left(\mathbf{Y_{\nu}^\dagger} L \mathbf{Y_{\nu}}\right)_{21} &\simeq&
  \frac{\D -L_3 \, M_3 \, m_2 }{\D \sqrt{2}}
  \left( c_{12} \, \sin \omega_{13} + s_{12} \, e^{-i \alpha/2} \, \cos
    \omega_{13} \, \sin \omega_{23} \right) \nonumber\\ 
  && \times \left( e^{i \alpha/2} \,
    c_{12} \, \cos \omega_{13}^* \, \sin \omega_{23}^* - s_{12} \,
    \sin \omega_{13}^* \right) ~.  
\end{eqnarray}
%%%%%%%%%%%%%%%%%%%%%%%%%%%%%%%%%%%%%%%%%%%%%%
%
This serves to underline that -- as in the case of a NH light
neutrino spectrum discussed in Section~\ref{sec:NHNH} -- we have
typically $\left|\left(\mathbf{Y_{\nu}^\dagger} L \mathbf{Y_{\nu}}\right)_{21}\right| \sim
M_3 \, \sqrt{\dma}/v_u^2$. This leads for 
$M_3\simeq(10^{14}-10^{15})$ GeV and $m_0$, $A_0$ and $m_{1/2}$ in the few$\times$100 GeV
range to a $\mu\to e + \gamma$ decay branching ratio which exceeds the
existing limit by approximately 3 orders of magnitude.  Looking again
for simplifications with interesting phenomenological consequences, we
can reduce the magnitude of $\left|\left(\mathbf{Y_{\nu}^\dagger} L
\mathbf{Y_{\nu}}\right)_{21}\right|$ by setting $\omega_{13} = \omega_{23} = 0$
and, correspondingly, $\mathbf{R}_{13} = \mathbf{R}_{23} = \mathbf{1}$
in eq.~(\ref{R3rot})~\footnote{If one uses a somewhat different
  parametrisation of $\mathbf{R}$, namely,
  $\mathbf{R}=\mathbf{R}_{12}\mathbf{R}_{23}\mathbf{R'}_{12}$, the
  same result is achieved by setting just $\omega_{23} = 0$.}.  The
corresponding form of $R$ is
%%%%%%%%%%%%%%%%%%%%%%%%%%%%%%%%%%%%%%%%%%%%%%%%%%%%
\begin{equation}
\mathbf{R} \simeq 
\left( 
\bad 
\cos \omega_{12} & \sin \omega_{12} & 0 \\
-\sin \omega_{12} & \cos \omega_{12} & 0 \\
0 & 0 & 1 
\ea \right)~. 
\label{RIH}
\end{equation}
%%%%%%%%%%%%%%%%%%%%%%%%%%%%%%%%%%%%%%%%%%%%%%%%%
%
With $\mathbf{R}$ given by eq.~(\ref{RIH}) and negligible
$m_3/m_{1,2}$, the heaviest (RH) Majorana neutrino $N_3$ decouples and
we have again $(\mathbf{Y_{\nu}})_{3j} = 0~(j=1,2,3)$.  Neglecting
further the splitting between $m_1$ and $m_2$ we find:
%%%%%%%%%%%%%%%%%%%%%%%%%%%%%%%%%%%%%%%%%%%%%%%%%%
\begin{eqnarray}
  \left(\mathbf{Y_{\nu}^\dagger} L \mathbf{Y_{\nu}}\right)_{21} \!&\simeq&\! 
  - \frac{\D L_2 \, M_2 \,\sqrt{|\deltaatm|}}{\D v_u^2} \, \left( c_{12}
    \, \sin \omega_{12} - e^{-i \alpha/2} \, s_{12} \, \cos \omega_{12}
  \right) \nonumber\\ %[0.3cm]
  && \times \left[ e^{i \alpha/2} \, c_{12} \, c_{23} \, \cos \omega_{12}^* +
    (s_{12} \, c_{23} + e^{i \delta} \, c_{12} \, s_{23} \, s_{13} ) \,
    \sin \omega_{12}^* \right]~,
  \label{YY21IH}
\end{eqnarray}
%%%%%%%%%%%%%%%%%%%%%%%%%%%%%%%%%%%%%%%%%%%%%%%%%%%%%%
%
where we have used $m_{1,2}\cong\sqrt{\left|\deltaatm\right|}$ and have
neglected terms $\propto s_{13}$ which give a correction not bigger
than approximately $13 \%$.  Being of the order 
$M_2 \sqrt{\left|\deltaatm\right|}/v_u^2$, the expression~(\ref{YY21IH}) for
$\left(\mathbf{Y_{\nu}^\dagger} L \mathbf{Y_{\nu}}\right)_{21}$ will lead for
$M_2 \simeq 10^{12}$ GeV and values of the soft SUSY breaking
parameters in the few$\times$100 GeV range to $\text{B}(\mu \to e +
\gamma)$ close to the existing limits.  For
$\left(\mathbf{Y_{\nu}^\dagger}L \mathbf{Y_{\nu}}\right)_{31,32}$ we similarly
get
%%%%%%%%%%%%%%%%%%%%%%%%%%%%%%%%%%%%%%%%%%%%%%%%
\begin{eqnarray}
\left(\mathbf{Y_{\nu}^\dagger} L \mathbf{Y_{\nu}}\right)_{31} \!&\simeq&\!
- \frac{\D L_2 \, M_2 \, \sqrt{\left|\deltaatm\right|} }{\D v_u^2} \, \left(
   e^{-i \alpha/2} \, s_{12} \, \cos \omega_{12} - c_{12} \, \sin
   \omega_{12}
\right) \nonumber\\ 
&& \times\left[ e^{i \alpha/2} \, c_{12} \, s_{23} \, \cos \omega_{12}^* +
   (s_{12} \, s_{23} - e^{i \delta} \, c_{12} \, c_{23} \, s_{13} ) \,
   \sin \omega_{12}^* \right] \,,
\label{YY31IH} \\[0.3cm]
%\end{eqnarray}
%%%%%%%%%%%%%%%%%%%%%%%%%%%%%%%%%%%%%%%%%%%%%%%%%%%
%and 
%%%%%%%%%%%%%%%%%%%%%%%%%%%%%%%%%%%%%%%%%%%%%%%%%%%
%\begin{eqnarray}
\left(\mathbf{Y_{\nu}^\dagger} L \mathbf{Y_{\nu}}\right)_{32} &\simeq& 
- \frac{\D L_2 M_2 \sqrt{\left|\deltaatm\right|} }{\D v_u^2} \, 
\left[ \left( c_{23}\,s_{12} + e^{-i\,\delta} \, c_{12}\,s_{23}\,
    s_{13} \right) \sin \omega_{12} + e^{-i\alpha/2}
   \,c_{12}\,c_{23}\,\cos \omega_{12}
\right] \nonumber\\
&&\times \left[ e^{ i\alpha/2}\, c_{12}\,s_{23}\,\cos\omega_{12}^* + \left(
     s_{12}\,s_{23} - e^{i\,\delta}\,c_{12}\, c_{23}\,s_{13} \right )
   \sin \omega_{12}^*\right]\;.
\label{YY32IH}
\end{eqnarray}
%%%%%%%%%%%%%%%%%%%%%%%%%%%%%%%%%%%%%%%%%%%%%%%%%%%%%%%%%%
%
We see that, as in the case of NH light neutrino mass spectrum,
$\left(\mathbf{Y_{\nu}^\dagger} L \mathbf{Y_{\nu}}\right)_{21}$ and
$\left(\mathbf{Y_{\nu}^\dagger} L \mathbf{Y_{\nu}}\right)_{31}$ are rather
similar in structure, whereas $\left(\mathbf{Y_{\nu}^\dagger} L
\mathbf{Y_{\nu}}\right)_{32}$ differs somewhat. The Majorana phase $\beta_M$
does not appear in the expressions~(\ref{YY21IH})--(\ref{YY32IH})
because we have set $m_3/m_{1,2}=0$.  As the phase factor including
the Dirac phase $\delta$ appears always multiplied by the small
parameter $s_{13}$, for $s_{13}<0.1$ the branching ratios depend
essentially only on the Majorana phase $\alpha$, which enters also
into the expression for the effective Majorana mass $\meff$ in
$\betabeta$-decay~\cite{BGKP96,BPP1,STPFocusNu04}:
%%%%%%%%%%%%%%%%%%%%%%%%%%%%%%%
\begin{equation}
\meff \cong \sqrt{\Delta m^2_{13}}
\left|\cos^2\theta_{12}  + 
e^{i\alpha}~\sin^2 \theta_{12} \right|\;.
\label{meffIH1}
\end{equation}
%%%%%%%%%%%%%%%%%%%%%%%%%%%%%%%%

We will give next the ratios of $\text{B}(\mu \to e + \gamma)$ and
$\text{B}(\tau \to e + \gamma)$ ($\text{B}(\tau \to \mu + \gamma)$) in
the case of negligible contribution of the terms $\propto
s_{13}$~\footnote{For $s_{13} < 0.10$ the correction due to the terms
  in question can be shown to be smaller than approximately 15\%.}:
%%%%%%%%%%%%%%%%%%%%%%%%%%%%%%%%%%%%%%%%%%%%%%%%%%%%%%%%%%
\begin{eqnarray}
  \label{R2131IH}
&& \text{R}(21/31) \simeq \cot^2\theta_{23}\,, \\
&&\text{R}(21/32) \simeq 
s^{-2}_{23}~\frac{\D \left | e^{i \alpha/2}\,
        c_{12}\,\sin \omega_{12} -
       s_{12} \, \cos \omega_{12} \right |^2 \,
     \left | c_{12}\,\sin \omega_{12} -e^{i \alpha/2}\,
          s_{12}\, \cos \omega_{12} 
           \right |^2 }
{\D \left | e^{i \alpha/2}\,s_{12} \, \sin \omega_{12} 
       + c_{12} \, \cos \omega_{12} \right |^2 \,
     \left | 
       s_{12}\,\sin \omega_{12} + e^{i \alpha/2}\,c_{12}\,
        \cos \omega_{12} \right |^2}\,. \label{R2132IH}
\end{eqnarray}
%%%%%%%%%%%%%%%%%%%%%%%%%%%%%%%%%%%%%%%%%%%%%%%%%%%%%
%
Hence, as in the case of NH light neutrino mass spectrum, $\text{R}(21/31)$
is rather close to one, whereas $\text{R}(21/32)$ can have a wide range of
values.  Most interestingly, $\text{R}(21/32)$ can have a value close to two
or even be as large as $\sim 10$.

%%%%%%%%%%%%%%%%%%%%%%%%%%%%%%%%%
%
\subsubsection{\label{sec:ihnh_YB}
\large{Leptogenesis Constraints}}
%
%%%%%%%%%%%%%%%%%%%%%%%%%%%%%%%%

\indent We can again work out possible constraints from the
requirement of successful leptogenesis.  Using expression~(\ref{RIH})
for the matrix $\mathbf{R}$ and eq.~(\ref{e1H}) we find that the
CP-violating decay asymmetry $\epsilon_1$ of interest has the form
%%%%%%%%%%%%%%%%%%%%%%%%%%%%%%%%%%%%%%%%%%%
\begin{eqnarray} 
\label{eq:epsIH1}
\epsilon_1 \!\!&\simeq&\!\! - ~\frac{3}{8\pi} 
\left( \frac{m_2~M_1}{v_u^2}\right)\,
\frac{\deltasol}{\left|\deltaatm\right|}\,
\frac{{\rm Im}\left[ \sin^2\omega_{12} \right] }
{\left( 1 + \frac{\deltasol}{ 2 \left|\deltaatm\right|}\right )\, 
\left|\sin\omega_{12} \right|^2 + \left|\cos\omega_{12}\right|^2} \nonumber\\
\!&\simeq&\! - ~\frac{3}{16\pi} \left (\frac{m_2\, M_1}{v_u^2}\right )
\frac{\deltasol}{\left|\deltaatm\right|}\,
\sin 2 \rho \, \tanh 2\sigma~, 
\label{eq:epsIH2}
\end{eqnarray}
%%%%%%%%%%%%%%%%%%%%%%%%%%%%%%%%%%%%%%%%%%%%%%
%
where $\omega_{12} = \rho+i\sigma$, $m_2 \cong \sqrt{|\deltaatm|}$ and
we have neglected corrections $\sim \deltasol/\left|\deltaatm\right|$.  We see
that in order to have $\epsilon_1 \neq 0$, both $\rho$ and $\sigma$
should be different from zero: $\rho \neq k\pi/2$, $k=0,1,2...$,
$\sigma \neq 0$. Since $\epsilon_1 < 0$, we should have $\sin 2
\rho\tanh 2\sigma > 0$.

It follows from eq.~(\ref{eq:epsIH1}) that in the case of IH light
neutrino mass spectrum under discussion, the CP-asymmetry $\epsilon_1$
is suppressed by the factor $\deltasol/\left|\deltaatm\right|$.  The expression
for the CP-asymmetry we have found for the NH spectrum, 
eq.~(\ref{eq:epsNH2}), does not contain the indicated suppression factor.
It is not difficult to show that one always has
%%%%%%%%%%%%%%%%%%%%%%%%%%%%%%%%%
\begin{equation}
 \frac{\left |{\rm Im}\left[ \sin^2\omega_{12} \right] \right|}
{\left ( 1 + \frac{\deltasol}{2 \left|\deltaatm\right|} \right)\,
\left|\sin\omega_{12}\right|^2
+ \left|\cos\omega_{12}\right|^2} \leq \frac{1}{2}\; .
\label{ulimIH}
\end{equation}
%%%%%%%%%%%%%%%%%%%%%%%%%%%%%%%
%
{}For the asymmetry $\epsilon_1$ we get the upper limit
%%%%%%%%%%%%%%%%%%%%%%%%%%%%%%%%%%
\begin{equation}
\left|\epsilon_1\right| \ltap \frac{3}{16\pi} \left (\frac{m_2\,M_1}{v_u^2}\right )
\frac{\deltasol}{|\deltaatm|}
\simeq 3.2\times 10^{-9} 
\left (\frac{m_2}{0.05~{\rm eV}}\right )\,
\left (\frac{M_1}{10^{9}~{\rm GeV}}\right )\,
\left (\frac{174~{\rm GeV}}{v_u}\right )^2\,, 
\label{maxasymIH}
\end{equation}
%%%%%%%%%%%%%%%%%%%%%%%%%%%%%%%%%%%%%%%%%%%
%
where we have used $\deltasol/\left|\deltaatm\right| = 3.2\times 10^{-2}$.  The
maximal value of $\epsilon_1$ is reached for $\rho = \pi/4$ and
$\sigma \gtap 0.5$.

{}For the neutrino mass parameter $\widetilde{m}_1$, eq.~(\ref{tilm1}), we
find:
%%%%%%%%%%%%%%%%%%%%%%%%%%%%%%%%%%%
\begin{align}
\widetilde{m}_1 \simeq m_{1,2}(|\cos\omega_{12}|^2+|\sin\omega_{12}|^2)
= m_{1,2}\cosh 2\sigma \geq m_{1,2}\;.
\label{tilm1IH}
\end{align}
%%%%%%%%%%%%%%%%%%%%%%%%%%%%%%%%%%%
%
%%%%
%%% Figure 3
%%%%
\begin{figure}[!t]
\begin{center}
\begin{tabular}{cc}
\includegraphics{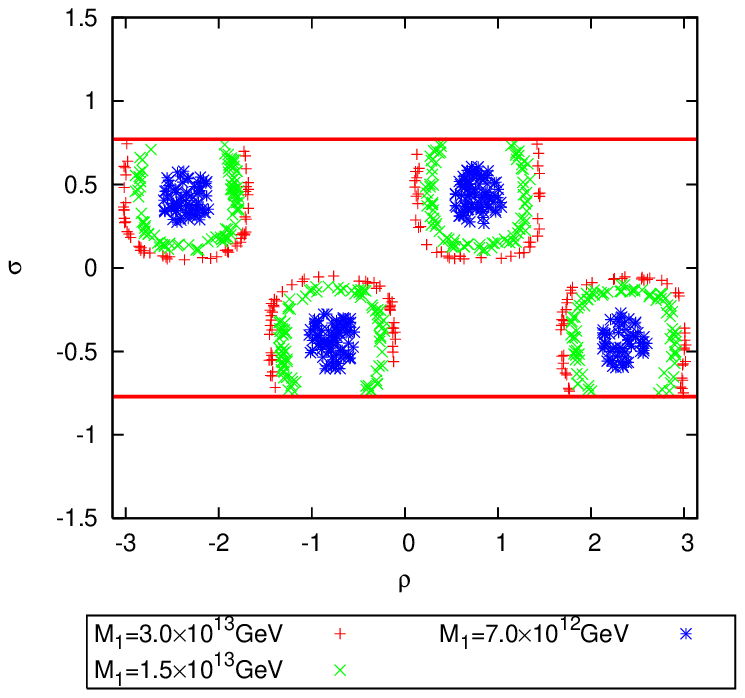}&
\includegraphics{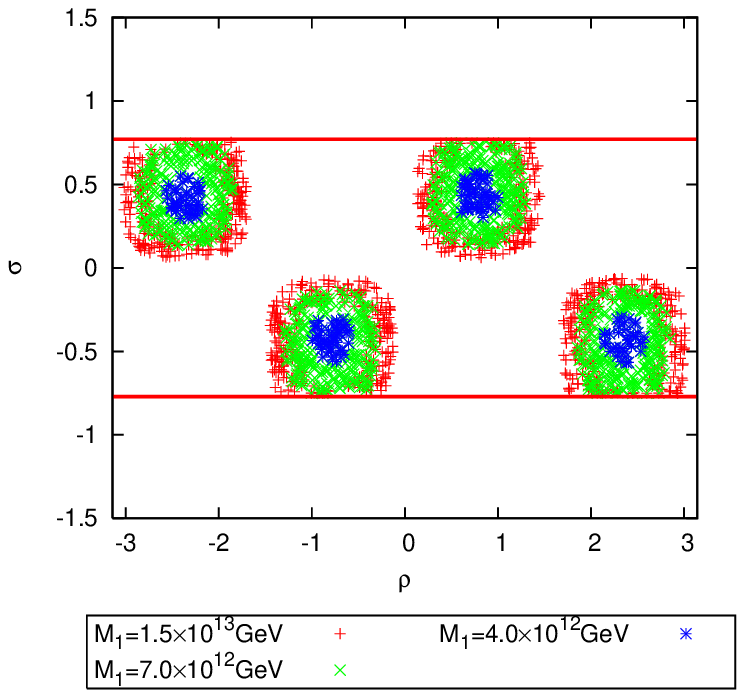}\\
(a)&(b)
\end{tabular}
%%%%%
\caption{The same as in Fig.~\ref{rho-sigma-nh},
but for IH light neutrino mass spectrum and
(a) $M_1 = 7\times 10^{12}$ GeV; $1.5\times 10^{13}$ GeV; $3\times 10^{13}$ GeV
(blue; green; red areas),
(b) $M_1 = 4\times 10^{12}$ GeV; $7.0\times 10^{12}$ GeV; $1.5\times 10^{13}$ GeV
(blue; green; red areas).
}
\label{rho-sigma-ih}
\end{center}
\end{figure}
%%%%%%%%%%%%%%%%%%%%%%%%%%%%%%%%%%%%%%%%%%%%%%%%%%%%%%%
%
The minimal value of $\widetilde{m}_1 = m_{1,2} \cong \sqrt{\left|\deltaatm\right|}
\cong 5\times 10^{-2}~{\rm eV}$, corresponds to $\cosh2\sigma = 1$ for
which $\left|\epsilon_1\right| = 0$.  For $5\times 10^{-2}~{\rm eV} < \widetilde{m}_1
\ltap 0.1$ eV, the efficiency factor lies in the interval $2.4\times
10^{-3} \ltap \kappa < 5.4\times 10^{-3}$.  It is not difficult to
convince oneself that for a given $M_1$, the maximal value of
$\left|\epsilon_1\right|$ is reached for $\sigma \cong \pm 0.5$, for which
$\widetilde{m}_1 \cong 7.8\times 10^{-2}~{\rm eV}$ and, correspondingly,
$\kappa \cong 3.2\times 10^{-3}$.  Thus, successful leptogenesis can
take place for $M_1 \gtap 6.7\times 10^{12}$ GeV, where we have used
eq.~(\ref{YBobs}).  In Fig.~\ref{rho-sigma-ih} we show the regions of
values of $\rho$ and $\sigma$, favoured by requirement of successful
thermal leptogenesis, for three fixed values of $M_1 = 7\times
10^{12}$ GeV; $1.5\times 10^{13}$ GeV; $3\times 10^{13}$ GeV.  We find that
$\left|\sigma\right| \ltap 0.75$, $|\sigma| \neq 0$.
%%%%%%%%%%%%%%%%%%%%%%%%%%%%%%%%%%%%%%%%%%%%%%%%%%%%%%%
%%%%
%%% Figure 4
%%%%
\begin{figure}[!t]
\begin{center}
\includegraphics{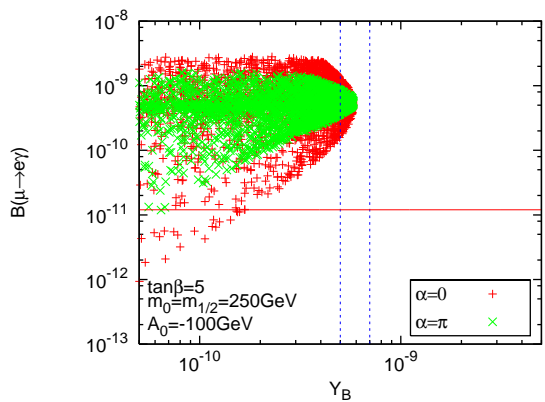}
%\end{center}
\caption{
  The correlation between the predicted baryon asymmetry of the
  Universe $Y_B$ and the predicted branching ratio of $\mu\to e +
  \gamma$ decay $\text{B}(\mu\to e+\gamma)$ in the case of IH light
  neutrino mass spectrum and for $M_1=7.0\times 10^{12}$ GeV and
  $M_2=4.0\times 10^{13}$ GeV.  The figure was obtained for the
  ``benchmark'' values of the soft SUSY breaking parameters $m_0 =
  m_{1/2} = 250$ GeV, $a_0m_0 = -100$ GeV and $\tan\beta = 5$.  The
  horizontal line indicates the experimental upper limit on
  $\text{B}(\mu\to e+\gamma)$, while the region between the two
  vertical dashed lines is favoured by the observed value of the baryon
  asymmetry of the Universe, $5.0\times 10^{-10}\leq Y_B\leq 7.0\times
  10^{-10}$.  }
\label{etaB-meg-ih}
\end{center}
%\vspace*{-1cm}
\end{figure}
%%%%%%%%%%%%%%%%%%%%%%%%%%%%%%%%%%%%%%%%%%%%%%%%%%%%%%%
%%%%
%%% Figure 5
%%%%
\begin{figure}[!!t]
\begin{center}
%%%%%
\begin{tabular}{cc}
\includegraphics{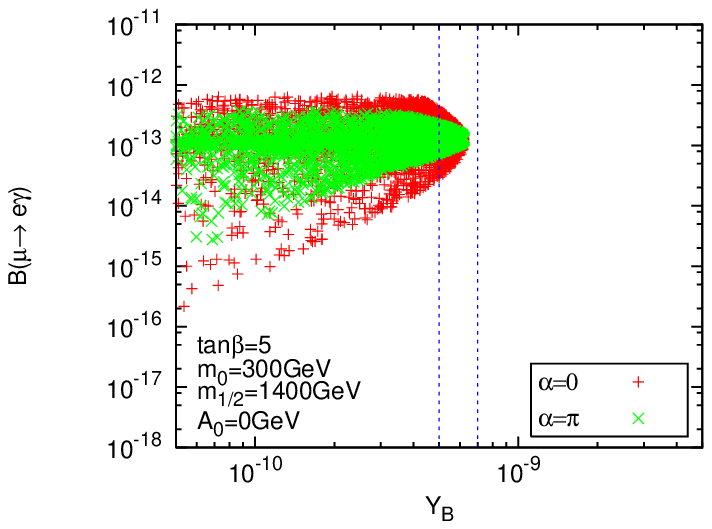}&
\includegraphics{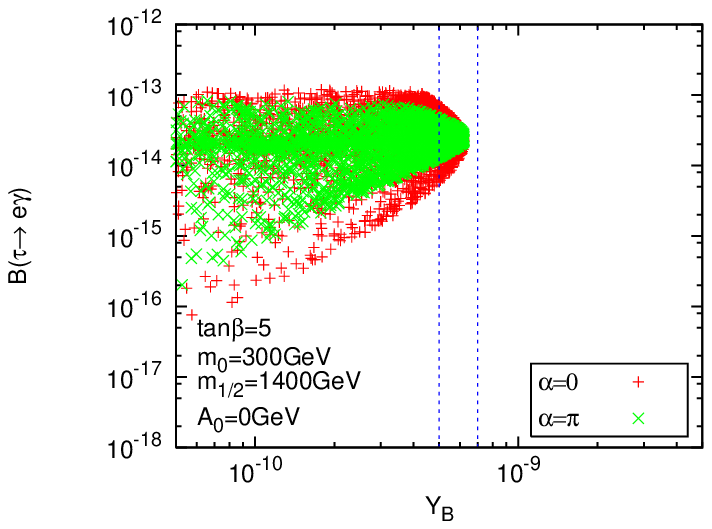}\\
(a)&(b)\\
\includegraphics{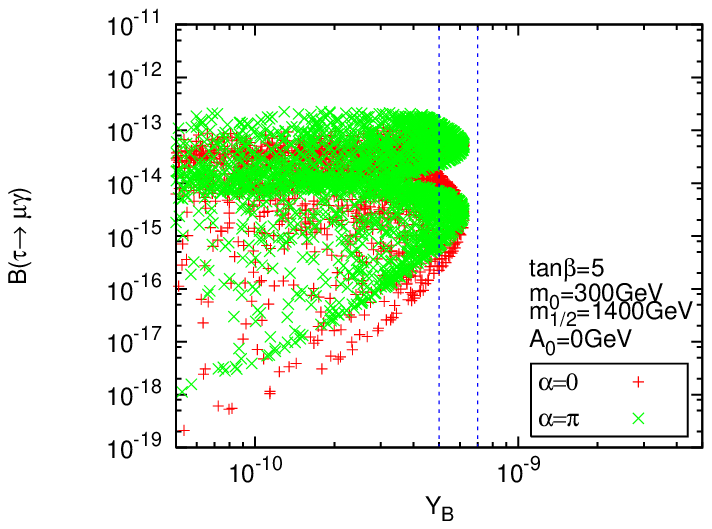}&\\
(c)&
\end{tabular}
%%%%%
\caption{The correlation between the 
predicted $Y_B$ and the predicted
$\text{B}(\mu\to e+\gamma)$ (a),
$\text{B}(\tau\to e+\gamma)$ (b),
$\text{B}(\tau\to \mu+\gamma)$ (c),
for IH light neutrino mass spectrum and
$M_1=7.0\times 10^{13}$ GeV, 
$M_2=4.0\times 10^{13}$ GeV.
Th results shown are obtained for
two values of the Majorana phase
$\alpha = 0;~\pi$ and
the following set of values of
the SUSY parameters:
$m_0=300$ GeV, $m_{1/2}=1400$ GeV, $a_0=0$ and
$\tan\beta=5$. 
The region between the two vertical dashed lines
corresponds to
$5.0\times 10^{-10}\leq Y_B\leq 7.0\times 10^{-10}$
and is favoured by the 
measured value of the baryon asymmetry.
The green (red+green) areas correspond to the 
Majorana phase $\alpha = \pi~(0)$.
}
\label{etaB-lfv-ih}
\end{center}
\end{figure}
%%%%%%%%%%%%%%%%%%%%%%%%%%%%%%%%%%%%%%%%%%%%%%%%%%%%%%%
Given the minimal value of $M_1$ determined by the leptogenesis
constraint, we will consider further in this subsection the following
hypothetical heavy Majorana neutrino mass spectrum: $M_1 = 7.0\times
10^{12}$ GeV, $M_2 = 4.0\times 10^{13}$ GeV, $M_3 = 2.0\times 10^{14}$
GeV.  For $M_2 = 4.0\times 10^{13}$ GeV, we find
$M_2\sqrt{\left|\deltaatm\right|} L_2/v_u^2\cong 0.4$.  At the same time the upper
limit on $\text{B}(\mu \to e + \gamma)$ implies that for the
``benchmark'' values of the soft SUSY breaking parameters we have
specified earlier and $\tan\beta=5$, we should have
$\left|\left(\mathbf{Y_{\nu}^\dagger} L \mathbf{Y_{\nu}}\right)_{21}\right|^2 \ltap
4.8\times 10^{-4}$.  It follows from the explicit expression for
$\left|\left(\mathbf{Y_{\nu}^\dagger} L \mathbf{Y_{\nu}}\right)_{21}\right|^2$,
eq.~(\ref{YY21IH}), that this upper limit is impossible to satisfy
for the values of the parameters $\rho$ and $\sigma$ satisfying the
leptogenesis constraint (Fig.~\ref{rho-sigma-ih}).  This is clearly
seen in Fig.~\ref{etaB-meg-ih}, which shows that the requirement of
successful leptogenesis and the existing experimental upper limit on
$\text{B}(\mu\to e + \gamma)$ are incompatible in the case of the
``benchmark'' values of the SUSY parameters, $m_0 = m_{1/2} = 250$
GeV, $a_0 m_0 = -100$ GeV, and of $\tan\beta = 5$.  The result we have
obtained indicates that in the case of IH light neutrino mass
spectrum, the SUSY parameters $m_0$ and/or $m_{1/2}$ should have
values considerably larger than the ``benchmark'' values we consider.
More specifically, we can have $m_0 \sim (250-300)$ GeV, but $ m^2_0
\ll m^2_{1/2}$.  This possibility is illustrated in
Fig.~\ref{etaB-lfv-ih}, where the predicted values of $\text{B}(l_i\to
l_j + \gamma)$ for $m_0=300$ GeV, $m_{1/2}=1400$ GeV, $a_0m_0=0$, and
$\tan\beta=5$, are shown as functions of the predicted value of the
baryon asymmetry.  The figure corresponds to $M_1 = 7\times 10^{12}$
GeV and $M_2 = 4.0\times 10^{13}$ GeV.  Now the requirement for
successful leptogenesis is compatible with the existing constraint on
$\text{B}(\mu\to e + \gamma)$: for the values of $\rho$ and $\sigma$
ensuring successful leptogenesis we find that typically $3\times
10^{-14} \ltap \text{B}(\mu\to e +\gamma)\ltap 5\times 10^{-13}$.
Significantly larger values of $\text{B}(\mu\to e + \gamma)$ are
possible if $\tan\beta \gtap 10$. As Fig.~\ref{etaB-lfv-ih} also
shows, the predicted branching ratios $\text{B}(l_i \to l_j +\gamma)$
exhibit weak dependence on the Majorana phase $\alpha$.

If the light neutrino mass spectrum is of the IH type, the results we
have obtained in this subsection can have important implications for
the predicted spectrum of SUSY particles in the few 100 GeV -- 1 TeV
region, to be probed by the experiments at the LHC.  For $m_0=300$ GeV,
$m_{1/2}=1400$ GeV, $a_0m_0=0$, and $\tan\beta=5$, the lightest SUSY
particle is still a neutralino and its mass is approximately $600$
GeV. The mass of next to the lightest SUSY particle, which is a stau,
is very close to the mass of the lightest neutralino.  At the same
time the squarks are predicted to be relatively heavy, having masses
$\sim(2-3)$ TeV.

%%%%%%%%%%%%%%%%%%%%%%%%%%%%%%%
\subsection{\label{sec:QDNH}
\large{Quasi-Degenerate Light Neutrinos}}
%%%%%%%%%%%%%%%%%%%%%%%%%%%%%%%
%
\indent\ In this case one has $m_1 \cong m_2 \cong m_3 \equiv m$, with
$m \gs 0.1$ eV. It is easy to see that $\left(\mathbf{Y_{\nu}^\dagger} L
\mathbf{Y_{\nu}}\right)_{21}$ will be proportional to $M_3 m/v_u^2$,
and therefore a too large branching ratio for $\mu \to e + \gamma$ decay
will be predicted.  Indeed, setting $M_1 =0$ and $M_2=0$ and using the
complex Euler angle parametrisation
$\mathbf{R}=\mathbf{R}_{12}^{\prime}\mathbf{R}_{23}\mathbf{R}_{12}$,
we find:
%%%%%%%%%%%%%%%%%%%%%%%%%%%%%%%%%%%%%%%%%%%%%%%%%
\begin{eqnarray}
\left(\mathbf{Y_{\nu}^\dagger} L 
\mathbf{Y_{\nu}}\right)_{21} \!&\simeq&\! \frac{L_3mM_3}{\sqrt{2}v_u^2}
\left[
-(s_{12}+e^{i\delta}s_{13}c_{12})\sin\omega_{23}^*\sin\omega_{12}^*
-e^{i\frac{\alpha}{2}}c_{12}\cos\omega_{12}^*\sin\omega_{23}^*
+e^{i\frac{\beta_{M}}{2}}c_{13}\cos\omega_{23}^*
\right]\nonumber\\
&& \times
\left[
c_{13}(c_{12}\sin\omega_{12}-e^{-i\frac{\alpha}{2}}s_{12}\cos\omega_{12})
\sin\omega_{23}
+e^{-\frac{i}{2}(\beta_{M}-2\delta)}s_{13}\cos\omega_{23}
\right]\;,
\label{YY21QD1}
\end{eqnarray}
%%%%%%%%%%%%%%%%%%%%%%%%%%%%%%%%%%%%%%%%%%%
%
where for simplicity we have set $\theta_{23}=\pi/4$ and have
neglected the sub-dominant terms $\propto s_{13}$.  There are two
possibilities for suppression of
$\left|\left(\mathbf{Y_{\nu}^\dagger}L\mathbf{Y_{\nu}}\right)_{21}\right|$:  $(i)$ If
$\sin\omega_{23}=0$, the contribution due to $M_3$ in
$\left(\mathbf{Y_{\nu}^{\dagger}} L \mathbf{Y_{\nu}}\right)_{21}$ remains, but is
proportional to $s_{13}$.  The necessary suppression can take place if
$s_{13}$ is sufficiently small.  $(ii)$ The parameters $\omega_{12}$ and
$\omega_{23}$ can have values such that the different terms $\propto
M_3$ in $\left(\mathbf{Y_{\nu}^{\dagger}} L\mathbf{Y_{\nu}}\right)_{21}$ cancel (completely
or partially) each other.  The latter seems to require fine tuning
between the values of several very different parameters.

In what follows we shall consider the case $(i)$ and we set
$\omega_{23}=0$. In this case $\mathbf{R}$ has the form given in
eq.~(\ref{RIH}).  The quantities of interest
$\left(\mathbf{Y_{\nu}^{\dagger}} L \mathbf{Y_{\nu}}\right)_{ij}~(i\neq j)$, including the
contributions $\propto M_2$, are given by:
%%%%%%%%%%%%%%%%%%%%%%%%%%%%%%%%%%%%%%%%%%%%%
\begin{eqnarray}
\left(\mathbf{Y_{\nu}^\dagger}L\mathbf{Y_{\nu}}\right)_{21}\!&=&\!
\frac{L_3mM_3}{v_u^2}e^{i\delta}s_{23}c_{13}s_{13}\nonumber\\
&&+\frac{L_2mM_2}{v_u^2}c_{13}
\left[
(c_{23}s_{12}+e^{i\delta}s_{23}c_{12}s_{13})s_{\omega}^*
+e^{i\frac{\alpha}{2}}c_{23}c_{12}c_{\omega}^*
\right]
(-c_{12}s_{\omega}+e^{-i\frac{\alpha}{2}}s_{12}c_{\omega})\;, \nonumber\\
&&\\ 
\label{YY21QD}
\left(\mathbf{Y_{\nu}^\dagger} L 
\mathbf{Y_{\nu}}\right)_{31}\!\!&=&\!\!
\frac{L_3mM_3}{v_u^2}e^{i\delta}c_{23}c_{13}s_{13}\nonumber\\
&&+\frac{L_2mM_2}{v_u^2}c_{13}
\left[
(-s_{23}s_{12}+e^{i\delta}c_{23}c_{12}s_{13})s_{\omega}^*
-e^{i\frac{\alpha}{2}}s_{23}c_{12}c_{\omega}^*
\right]
(-c_{12}s_{\omega}+e^{-i\frac{\alpha}{2}}s_{12}c_{\omega})\;, \nonumber \\
&& \\
\label{YY31QD}
\left(\mathbf{Y_{\nu}^\dagger} L 
\mathbf{Y_{\nu}}\right)_{32}\!\!&=&\!\!
\frac{L_3mM_3}{v_u^2}c_{13}^2c_{23}s_{23}
\nonumber\\
&&+\frac{L_2mM_2}{v_u^2}c_{13}
\left[
(-s_{23}s_{12}+e^{i\delta}c_{23}c_{12}s_{13})s_{\omega}^*
-e^{i\frac{\alpha}{2}}s_{23}c_{12}c_{\omega}^*
\right]\nonumber\\
&&\phantom{Space}\times
\left[
(c_{23}s_{12}+e^{-i\delta}s_{23}c_{12}s_{13})s_{\omega}
+e^{-i\frac{\alpha}{2}}c_{23}c_{12}c_{\omega}
\right]\;,
\label{YY32QD}
\end{eqnarray}
%%%%%%%%%%%%%%%%%%%%%%%%%%%%%%%%%%%%%%%%%%%%%%%%%
%
where $\omega=\omega_{12}+\omega_{12}^{\prime}$.  It is interesting to
note that the quantity $\left|\left(\mathbf{Y_{\nu}^\dagger}L
    \mathbf{Y_{\nu}}\right)_{32}\right|^2$, and correspondingly
$\text{B}(\tau \to \mu + \gamma)$, is not suppressed by the factor
$s_{13}^2$.  The effective Majorana mass in $\betabeta$-decay depends
in the case of QD spectrum on the CP-violation Majorana phase 
$\alpha$~\cite{BPP1,STPFocusNu04}, present in the 
expressions~(\ref{YY21QD})--(\ref{YY32QD}) for $\left(\mathbf{Y_{\nu}^\dagger} L
\mathbf{Y_{\nu}}\right)_{ij}~(i\neq j)$: 
$\meff\cong m \left|\cos^2\theta_{12}+e^{i\alpha}~\sin^2\theta_{12}\right|$.
%%%%%%%%%%%%%%%%%%%%%%%%%%%%%%%%%%%%%%%%%%%%%%%%%%%%%%%
%%%%
%%% Figure 6
%%%%
\begin{figure}[!t]
\begin{center}
\begin{tabular}{cc}
\includegraphics{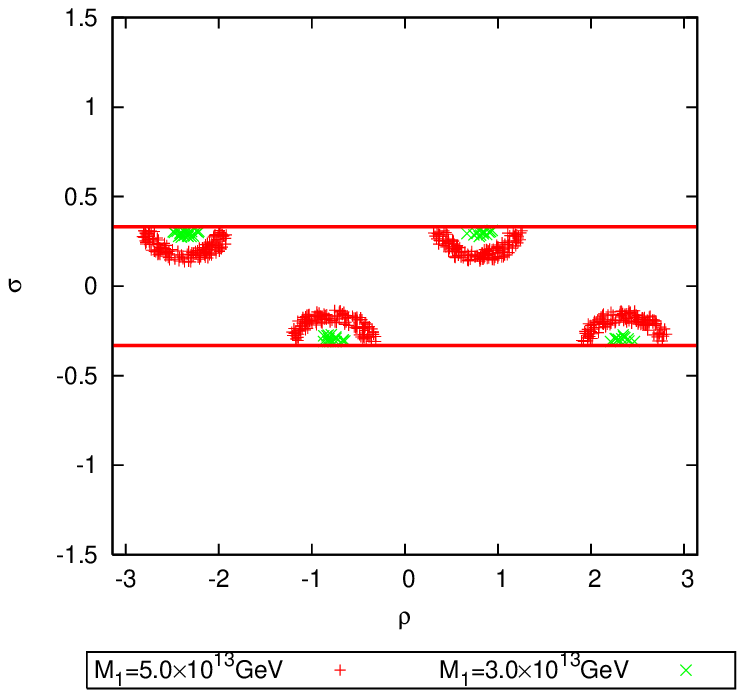}&
\includegraphics{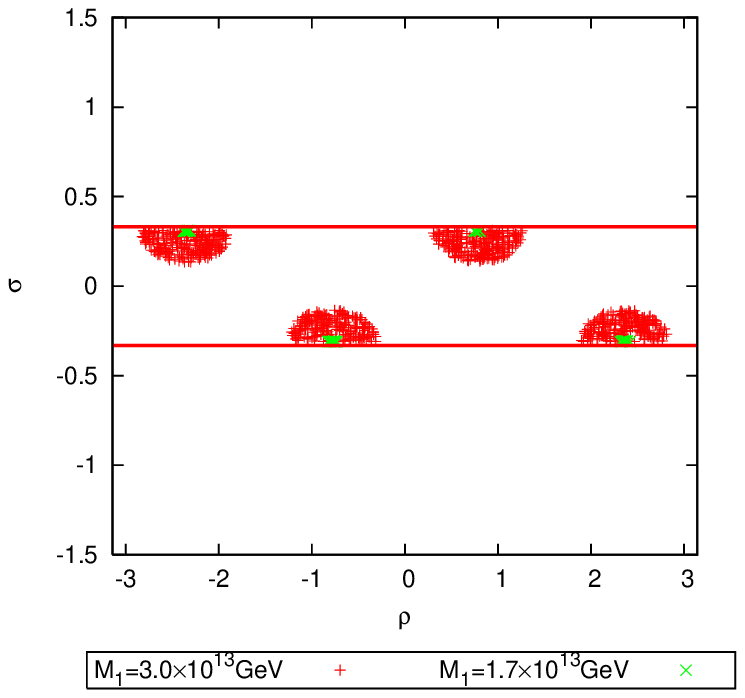}\\
(a)&(b)
\end{tabular}
%%%%%
\caption{The same as in Fig.~\ref{rho-sigma-nh},
but for QD light neutrino mass spectrum and
(a) $M_1 = 3.0\times 10^{13}$ GeV; $5.0\times 10^{13}$ GeV
(green; red areas),
(b) $M_1 = 1.7\times 10^{13}$ GeV; $3.0\times 10^{13}$ GeV
(green; red areas).}
\label{rho-sigma-qd}
\end{center}
\end{figure}

%%%%%%%%%%%%%%%%%%%%%%%%%%%%%%%%%%%%%%%%%
%
\subsubsection{Leptogenesis Constraints}
%
%%%%%%%%%%%%%%%%%%%%%%%%%%%%%%%%%%%%%%%
%
\indent\ For the CP-violating decay asymmetry $\epsilon_1$ we find
%%%%%%%%%%%%%%%%%%%%%%%%%%%%%%%%%%%%
\begin{align}
\epsilon_1
=& 
-\frac{3}{8 \pi}\left(\frac{m M_1}{v_u^2}\right)
\frac{\deltasol}{m^2}
\frac{\mathrm{Im}\left[s_{\omega}^2
\right]}
{
\left|c_{\omega}\right|^2
+\left(1+\frac{\Delta m_{\odot}^2}{2m^2}\right) \left|s_{\omega}\right|^2}\;.
\end{align}
%%%%%%%%%%%%%%%%%%%%%%%%%%%%%%%%%%%%
%
It is not difficult to show that
%%%%%%%%%%%%%%%%%%%%%%%%%%%%%%%%%%%%
\begin{align}
\frac{\mathrm{Im}(s_{\omega}^2)}
{\left|c_{\omega}\right|^2
+\left(1+\frac{\Delta m_{\odot}^2}{2m^2}\right) \left|s_{\omega}\right|^2
}
\simeq
\frac{\mathrm{Im}(s_{\omega}^2)}
{\left|c_{\omega}\right|^2 + \left| s_{\omega} \right|^2}
= \frac{1}{2}\sin2\rho\tanh 2\sigma\leq \frac{1}{2}\;.
\end{align}
%%%%%%%%%%%%%%%%%%%%%%%%%%%%%%%%%%%%%
%
Thus, the maximal asymmetry $\left|\epsilon_1\right|$ is given by
%%%%%%%%%%%%%%%%%%%%%%%%%%%%%%%%%%%%%
\begin{align}
\left|\epsilon_1\right|\leq&
1.6\times 10^{-9}
\left(\frac{0.1~\text{eV}}{m}\right)
\left(\frac{M_1}{10^9~\text{GeV}}\right)
\left(\frac{174~\text{GeV}}{v_u}\right)^2\;.
\label{maxasymQD}
\end{align}
%%%%%%%%%%%%%%%%%%%%%%%%%%%%%%%%%%%%
%
One can easily find also the mass parameter $\widetilde{m}_1$:
%%%%%%%%%%%%%%%%%%%%%%%%%%%%%%%%%%%%
\begin{align}
\widetilde{m}_1
\cong m \left[\left|c_{\omega}\right|^2+\left|s_{\omega}\right|^2\right]
= m \cosh 2\sigma\geq m\;.
\label{tilm1QD}
\end{align}
%%%%%%%%%%%%%%%%%%%%%%%%%%%%%%%%%%%%
%
%%%%%%%%%%%%%%%%%%%%%%%%%%%%%%%%%%%%%%%%%%%%%%%%%%%%%%%
%
%%%%
%%% Figure 7
%%%%
\begin{figure}[!t]
\begin{center}
\begin{center}
\includegraphics{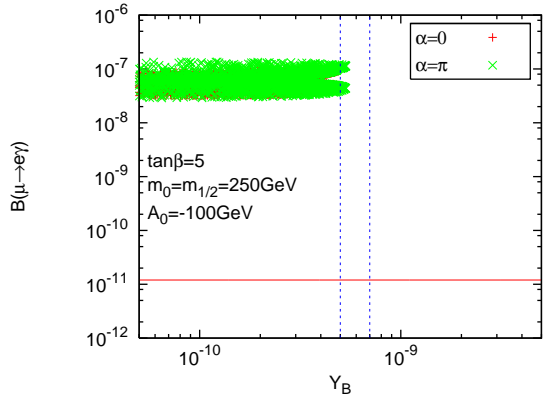}
\end{center}
\caption{
  The correlation between the predicted $Y_B$ and the predicted
  $\text{B}(\mu\to e+\gamma)$ in the case of QD light neutrino mass
  spectrum and for $M_1=3.0\times 10^{13}$ GeV, $M_2=1.2\times
  10^{14}$ GeV and $M_3=4.8\times 10^{14}$ GeV.  The ``benchmark''
  values of the soft SUSY breaking parameters $m_0 = m_{1/2} = 250$
  GeV, $a_0m_0 = -100$ GeV and $\tan\beta = 5$, have been used.  The
  horizontal line indicates the experimental upper limit on
  $\text{B}(\mu\to e+\gamma)$, while the region between the two
  vertical dashed lines corresponds to $5.0\times 10^{-10}\leq Y_B\leq
  7.0\times 10^{-10}$ and is favoured by the observed value of $Y_B$.
}
\label{etaB-meg-qd}
\end{center}
\end{figure}
%%%%%%%%%%%%%%%%%%%%%%%%%%%%%%%%%%%%%%%%%%%%%%%%%%%%%%%
%
%%%%%%%%%%%%%%%%%%%%%%%%%%%%%%%%%%%%%%%%%%%%%%%%%%%%%%%
%%%%
%%% Figure 8
%%%%
\begin{figure}[!t]
\begin{center}
%%%%%
\begin{tabular}{cc}
\includegraphics{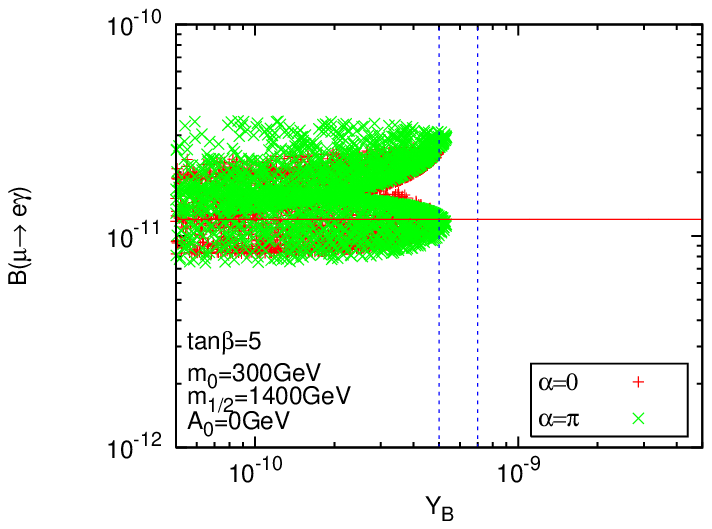}&
\includegraphics{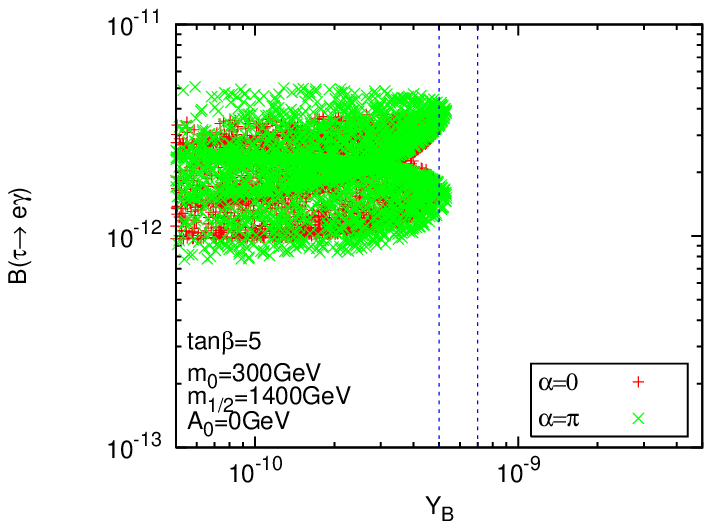}\\
(a)&(b)\\
\includegraphics{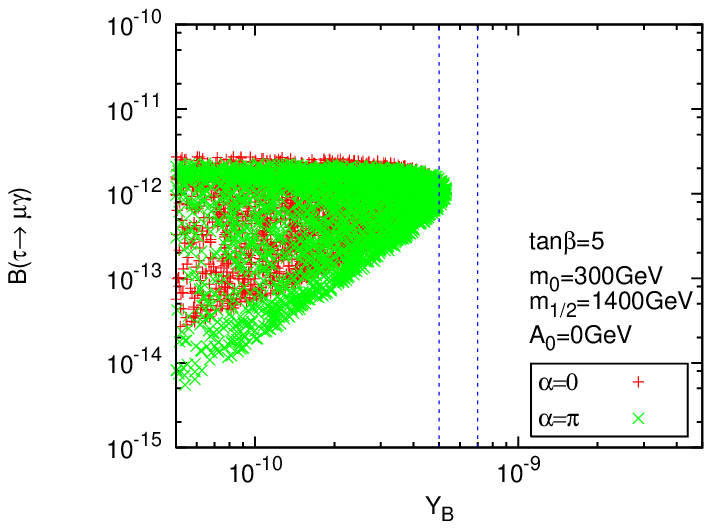}&\\
(c)&
\end{tabular}
%%%%%
\caption{
  The correlation between the predicted $Y_B$ and the predicted
  $\text{B}(\mu\to e+\gamma)$ (a), $\text{B}(\tau\to e+\gamma)$ (b),
  $\text{B}(\tau\to \mu+\gamma)$ (c), for QD light neutrino mass
  spectrum and $M_1=3.0\times 10^{13}$ GeV, $M_2=1.2\times 10^{14}$
  GeV, $M_2=4.8\times 10^{14}$ GeV.  The SUSY parameters used to
  obtain the figure are $m_0=300$ GeV, $m_{1/2}=1400$ GeV, $a_0=0$ and
  $\tan\beta=5$. Results for two values of the Majorana phase $\alpha
  = 0;~\pi$ are shown.  The region between the two vertical dashed
  lines corresponds to $5.0\times 10^{-10}\leq Y_B\leq 7.0\times
  10^{-10}$.  }
\label{etaB-lfv-qd}
\end{center}
\end{figure}
%
%%%%%%%%%%%%%%%%%%%%%%%%%%%%%%%%%%%%%%%%%%%%%%%%%%%%%%%
%
Since successful leptogenesis is possible 
for \cite{CERN04} $\widetilde{m}_1 \leq 0.12$ eV,
while for QD light neutrino mass spectrum $m\gtap 0.1$ eV,
we get from  eq. (\ref{tilm1QD}) that 
$m \cong 0.1$ eV. Therefore in all 
further analysis and numerical calculations 
in this subsection we set $m = 0.1$ eV.  

%  Given the fact that 
% for QD light neutrino mass spectrum $m\gtap 0.1$ eV, 
As it follows from the preceding discussion,
we have $\widetilde{m}_1 \cong (0.10 - 0.12)$ eV.
Correspondingly, the wash-out effect
in the case under consideration is relatively strong. 
Taking into account the precise upper limit 
on $\widetilde{m}_1$ given in 
ref.~\cite{CERN04}, $\widetilde{m}_1 \leq 0.12$ eV, we get for the
corresponding efficiency factor $1.9\times 10^{-3}\ltap \kappa \ltap
2.4\times 10^{-3}$.  The condition $\widetilde{m}_1 \leq 0.12$ eV implies
$\sigma \ltap 0.3$, for which $\tanh 2\sigma\ltap 0.5$. Thus, using
eqs.~(\ref{YBobsth}) and~(\ref{maxasymQD}) we obtain the minimal value
of $M_1$ ensuring successful leptogenesis: $M_1 \gtap 3.0\times
10^{13}$ GeV. Since $\widetilde{m}_1 \gtap 0.1$ eV, we can expect that
$\sigma$ lies in the interval $\sigma \cong (0.2 - 0.3)$.  This is
confirmed by a more detailed numerical analysis.  The values of the
parameters $\rho$ and $\sigma$ allowed by the leptogenesis constraint
are shown in Fig.~\ref{rho-sigma-qd} for $M_1 = 3.0\times 10^{13}$ GeV; 
$5.0\times 10^{13}$ GeV.

Given the lower bound $M_1 \gtap 3.0\times 10^{13}$ GeV, a possible
mildly hierarchical heavy Majorana neutrino mass spectrum would
correspond to, $\eg$, $M_1 = 3.0\times 10^{13}$ GeV, $M_2 = 1.2\times
10^{14}$ GeV, and $M_3 = 4.8\times 10^{14}$ GeV.  For this spectrum,
$L_3 m M_3/v_u^2\cong 6.0$ and $L_2 m M_2/v_u^2\cong 2.0$.  Using the
lowest possible value for $\tan^2\beta \cong 10$ we find that even if
$s_{13} = 0$ and the term $\propto M_3$ does not contribute to
$\left|\left(\mathbf{Y_{\nu}^\dagger} L \mathbf{Y_{\nu}}\right)_{21}\right|$, 
the contribution of the terms $\propto M_2$ is so large in the case of the
``benchmark'' values of the soft SUSY breaking parameters, $m_0=
m_{1/2} = 250$ GeV, $a_0m_0 = -100$ GeV, that the predicted
$\text{B}(\mu \to e + \gamma)$ exceeds the existing upper limit on
$\text{B}(\mu \to e + \gamma)$ by more than a factor of $\sim 10^{3}$.
The indicated incompatibility between the leptogenesis and
$\text{B}(\mu \to e + \gamma)$ constraints is illustrated 
in Fig.~\ref{etaB-meg-qd}.

Similarly to the case of IH light neutrino mass spectrum, the
requirement of successful thermal leptogenesis and the upper limit on
$\text{B}(\mu \to e + \gamma)$ can be simultaneously satisfied only
if the scale of masses of supersymmetric particles is significantly
larger than that predicted for the ``benchmark'' values of the soft
SUSY breaking parameters we have adopted.  In Fig.~\ref{etaB-lfv-qd}
we show the correlation between the predicted values of
$\text{B}(l_i\to l_j+ \gamma)$ for $\sin\theta_{13} = 0.05$ and
$\tan\beta=5$, $m_0 = 300$ GeV, $m_{1/2}=1400$ GeV and $a_0 = 0$, and
the predicted value of the baryon asymmetry.  The results presented in
this figure have been obtained for the spectrum of the heavy Majorana
neutrino masses specified above. We note, in particular, that the
predicted interval of values of $\text{B}(\mu\to e+\gamma)$ which is
compatible with the observed baryon asymmetry is in the range of
sensitivity of the ongoing MEG experiment: $\text{B}(\mu\to e+\gamma)$
can have a value just below the present experimental upper limit.  As
in the cases of NH and IH light neutrino mass spectra, we find that
the dependence of $\text{B}(l_i \to l_j+\gamma)$ on the relevant
Majorana phase $\alpha$ is rather weak.

%%% ADDED by T. SHINDOU%%%%%%%%%%%%%%%%%%%%%%%
\begin{figure}[!t]
\begin{tabular}{cc}
\includegraphics{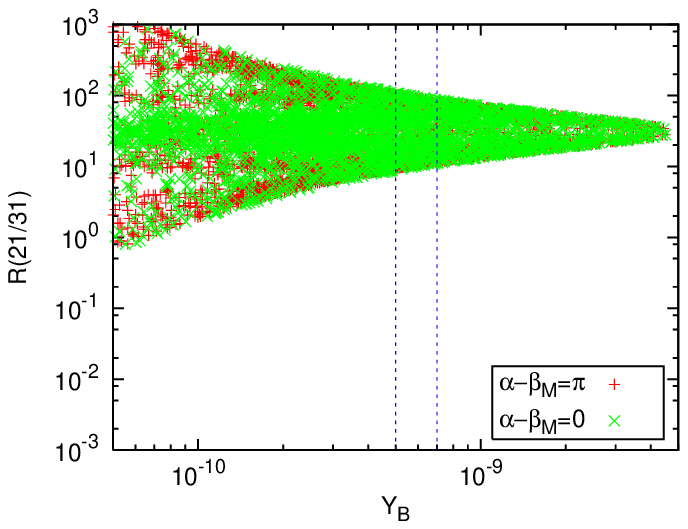}&
\includegraphics{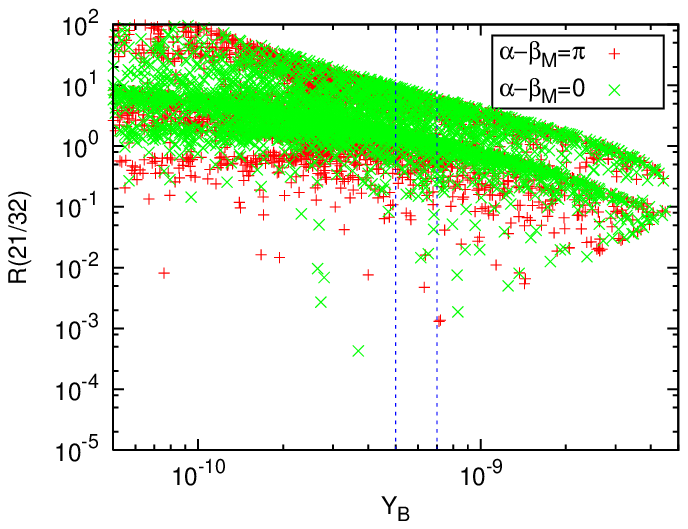}\\
\multicolumn{2}{c}{(NH)}\\
\includegraphics{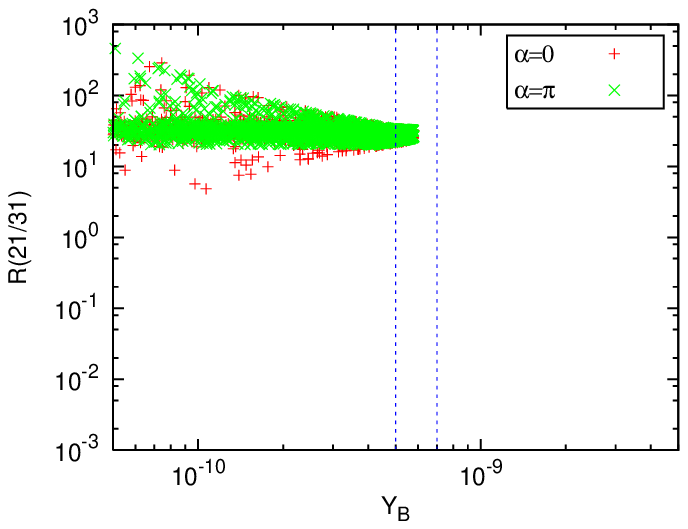}&
\includegraphics{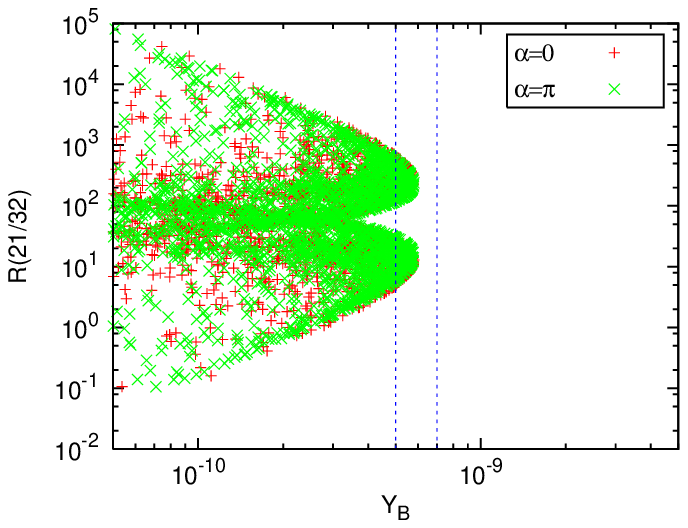}\\
\multicolumn{2}{c}{(IH)}\\
\includegraphics{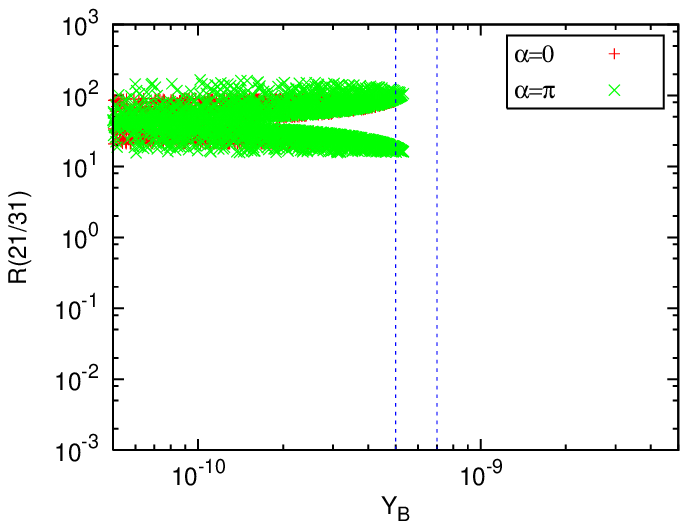}&
\includegraphics{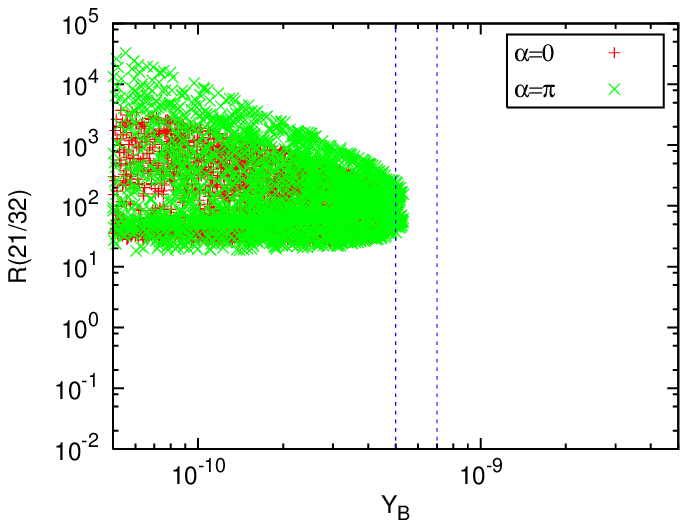}\\
\multicolumn{2}{c}{(QD)}
\end{tabular}
\caption{The correlation between the predicted $Y_B$ and 
the double ratios $\text{R}(21/31)$ and $\text{R}(21/32)$, defined in eq.~(\ref{DoubleR}).
The heavy neutrino mass spectrum in the NH, IH, and QD 
cases is taken to be same as in 
Fig.~\ref{etaB-lfv-nh}, Fig.~\ref{etaB-lfv-ih}
and Fig.~\ref{etaB-lfv-qd}, respectively.
The region between the two vertical dashed lines corresponds to 
$5.0\times 10^{-10}\leq Y_B\leq 7.0\times 10^{-10}$.
}
\label{ratio-BR}
\end{figure}
%%%%%%%%%%%%%%%%%%%%%%%%%%%%%%%%%%%%%%%%%%%%%%%%%%%
%
In Fig.~\ref{ratio-BR}, the correlations between the predicted value of
$Y_B$ and those of the double ratios $\text{R}(21/31)$ and $\text{R}(21/32)$ are displayed.
In the approximation we use, the double ratios are 
independent of SUSY parameters and are determined only by the
off-diagonal elements of $\mathbf{Y_{\nu}^{\dagger}}\mathbf{Y_{\nu}}$.
When the constraint of successful leptogenesis is imposed,
we get for the allowed range of values of $\text{R}(21/31)$ 
for the NH, IH and QD light neutrino mass spectrum respectively
$10\ltap \text{R}(21/31)\ltap 100$, $20\ltap \text{R}(21/31)\ltap 50$ 
and $\text{R}(21/31)\simeq 20;~100$. Similarly, for the double ratio
$\text{R}(21/32)$ we get in the three cases
$10^{-2}\ltap \text{R}(21/32)\ltap10$,
$10\ltap \text{R}(21/32)\ltap 10^3$ and 
$50\ltap \text{R}(21/32)\ltap 10^3$, respectively.
We find, in particular, that $\text{R}(21/32)$ can be 
much smaller than 1 only for NH light 
neutrino mass spectrum.

%%%%%%%%%%%%%%%%%%%%%%%%%%%%%%%%%%%%%%%%%%%%
%
\section{\large{Conclusions}}
%
%%%%%%%%%%%%%%%%%%%%%%%%%%%%%%%%%%%%%%%%%%%%

\indent\ We have considered the LFV decays $\mu \to e + \gamma$, $\tau
\to e + \gamma$ and $\tau \to \mu + \gamma$ and leptogenesis in the
MSSM with see-saw mechanism of neutrino mass generation and soft SUSY
breaking with universal boundary conditions at a scale $M_{\rm X} >
M_{\rm R}$, $M_{\rm R}$ being the heavy (RH) Majorana neutrino mass
scale.  The heavy Majorana neutrinos were assumed to have hierarchical
mass spectrum, $M_1\ll M_2\ll M_3$, while the scale $M_{\rm X}$ was
taken to be the GUT scale, $M_{\rm X} = 2\times 10^{16}$ GeV.  We have
derived the combined constraints, which the existing stringent upper
limit on the $\mu \to e + \gamma$ decay rate and the requirement of
successful leptogenesis impose on the neutrino Yukawa couplings, heavy
Majorana neutrino masses and SUSY parameters in the cases of the three
types of light neutrino mass spectrum -- normal and inverted
hierarchical (NH and IH), and quasi-degenerate (QD).  A basic quantity
in these analyses is the matrix of neutrino Yukawa couplings,
$\mathbf{Y_{\nu}}$.  In the present work we have used the orthogonal
parametrisation of $\mathbf{Y_{\nu}}$, in which $\mathbf{Y_{\nu}}$ is
expressed in terms of the light neutrino and heavy RH neutrino masses,
the PMNS neutrino mixing matrix $\pmns$, and an orthogonal matrix
$\mathbf{R}$.  Leptogenesis can take place only if $\mathbf{R}$ is
complex.

The constraints from thermal leptogenesis require, in general, that
$M_1 \gs 10^9$ GeV.  This would indicate a hierarchy, $\eg$, of the
form $M_1\simeq(10^{9}-10^{11})$ GeV, $M_2\simeq(10^{12}-10^{13})$ GeV
and $M_3\gtap 10^{13}~{\rm GeV}\gg M_2$, $M_3<(\ll)M_X$.  In our
analysis we have considered a ``benchmark SUSY scenario'' defined by
the values of the soft SUSY breaking parameters in the range of
few$\times$100 GeV: $m_0=m_{1/2}=250$ GeV, $a_0 m_0=-100$ GeV, and
$\tan\beta\sim(5-10)$.  In this scenario the lightest supersymmetric
particle is a neutralino with a mass of $\sim 100$ GeV.  The next to
the lightest SUSY particles are the chargino and a second neutralino
with masses $\sim 200$ GeV. The squarks have masses in the range of
$\sim (400-600)$ GeV.  Using the indicated set of ``benchmark'' values
of the soft SUSY breaking parameters and barring accidental
cancellations, we find that for the typical values of the heaviest
Majorana neutrino mass $M_3\cong(5\times10^{13}-10^{15})$ GeV, the
limit on the $\mu \to e + \gamma$ decay branching ratio $\text{B}(\mu
\to e + \gamma)$ is impossible to respect independently of the type of
the light neutrino mass spectrum: the predicted $\text{B}(\mu \to e +
\gamma)$ exceeds the existing upper limit by few orders of magnitude.
For each of the three types of neutrino mass spectrum -- NH, IH and
QD, we find simple forms of the matrix $\mathbf{R}$ which lead to a
suppression of the dominant contributions due to the terms $\propto
M_3$ in $\text{B}(\mu \to e + \gamma)$.  In all three cases the matrix
$\mathbf{R}$ ensuring the requisite suppression admits a
parametrisation by one complex angle.  In the case of NH spectrum
$\mathbf{R}$ is given by eq.~(\ref{RNH}), while for IH and QD spectra,
$\mathbf{R} \cong \mathbf{R}_{12}$, $\mathbf{R}_{12}$ being the matrix
of (complex) rotations in the 1-2 plane. In this case the dominant
contribution in $\text{B}(\mu \to e + \gamma)$ comes from terms
$\propto M_2$.  For QD spectrum the terms $\propto M_3$ are suppressed
by the factor $\sin\theta_{13}$ and can be comparable to those
$\propto M_2$.

The requirement of successful leptogenesis leads to a rather stringent
constraint on the complex mixing angle in $\mathbf{R}$. For IH and QD
spectra it also implies a relatively large lower limit on the mass of
the lightest RH Majorana neutrino: $M_1\gtap 7.0\times 10^{12}$ GeV
and $M_1\gtap 3.0\times 10^{13}$ GeV, respectively.  With such values
of $M_1$ and hierarchical heavy Majorana neutrino mass spectrum, the
upper bound on $\text{B}(\mu \to e + \gamma)$ can be satisfied only if
the scale of masses of SUSY particles is considerably higher than that
implied by the ``benchmark'' values of the soft SUSY breaking
parameters we have considered.  We have analysed a specific case of
such SUSY scenario: $m_0 = 300$ GeV, $m_{1/2} = 1400$ GeV and $a_0 =
0$.  In this scenario the lightest SUSY particle is a neutralino with
a mass of approximately $600$ GeV, the next to the lightest SUSY
particle is a stau and its mass is very close to the mass of the
lightest neutralino, while the squarks are relatively heavy, having
masses $\sim(2-3)$ TeV.  
The predictions for $\text{B}(\mu \to e + \gamma)$
are now largely in the range of
sensitivity of the ongoing MEG experiment.
If more stringent upper limits on
$\text{B}(\mu \to e + \gamma)$ will be obtained in the future, it
would be rather difficult to reconcile the IH and QD light neutrino
mass spectra with the $\mu \to e + \gamma$ and leptogenesis
constraints and SUSY particle masses in the TeV range.  Our
results may have important implications for the 
search of SUSY particles in the few$\times$100 GeV -- 1 TeV region,
to be performed by the experiments at the LHC.

Satisfying the combined constraints from the existing upper limit on
the $\mu \to e + \gamma$ decay rate and the requirement of successful
thermal leptogenesis proves to be a powerful tool to test the
viability of supersymmetric theories with see-saw mechanism of
neutrino mass generation and soft flavour-universal SUSY breaking at a
scale above the heavy RH Majorana neutrino mass scale.

\vspace{1cm} 
{\bf Acknowledgements.}  We would like to thank
P.~Di~Bari and M.~Raidal for useful correspondence.  This work was
supported in part by the Italian MIUR and INFN under the programs
``Fisica Astroparticellare'' (S.T.P. and T.S.).  The work of W.R. was
supported by the ``Deutsche Forschungsgemeinschaft'' in the
``Sonderforschungsbereich 375 f\"ur Astroteilchenphysik'' and under
project number RO-2516/3-1.

%\newpage

%

\begin{thebibliography}{99} 
\bibitem{sol}
B.~T.~Cleveland {\it et al.},
{Astrophys.\ J.}  {\bf 496} (1998) 505;
%%CITATION = ASJOA,496,505;%%
%
Y.~Fukuda {\it et al.} [Kamiokande Collaboration],
  {Phys.\ Rev.\ Lett.}  {\bf 77} (1996) 1683;
J.~N.~Abdurashitov {\it et al.}  [SAGE Collaboration],
{J.\ Exp.\ Theor.\ Phys.}  {\bf 95} (2002) 181;
%%CITATION = ASTRO-PH 0204245;%%
%
T.~Kirsten {\it et al.} [GALLEX and GNO Collaborations], 
{Nucl.\ Phys.\ B (Proc. Suppl.)}  {\bf 118} (2003) 33;
C.~Cattadori {\it et al.},
{Nucl.\ Phys.\ B (Proc. Suppl.)}
{\bf 143} (2005) 3.
%
\bibitem{SKsolaratm}
S.~Fukuda {\it et al.} [Super-Kamiokande Collaboration],
{Phys.\ Lett.} {\bf B539} (2002) 179;
%
% Y.~Fukuda {\it et al.} 
% {Phys.\ Rev.\ Lett.}  {\bf 81} (1998) 1562;
%%CITATION = HEP-EX 9807003;%%
%
Y.~Ashie {\it et al.}, 
{Phys.\ Rev.\ Lett.}  {\bf 93} (2004) 101801; 
{Phys.\ Rev.} {\bf D71} (2005) 112005. 
% hep-ex/0501064
%%CITATION = HEP-EX 0404034;%%
%%CITATION = HEP-EX 0501064;%%
%
\bibitem{SNO123}
Q.~R.~Ahmad {\it et al.}  [SNO Collaboration],
{Phys.\ Rev.\ Lett.}  {\bf 87} (2001) 071301;
{\it ibid.} {\bf 89} (2002) 011301; 
{\it ibid.} {\bf 89} (2002) 011302;
S.~N.~Ahmed {\it et al.},
{Phys.\ Rev.\ Lett.}  {\bf 92} (2004) 181301;
B.~Aharmim {\it et al.}, nucl-ex/0502021.
%%CITATION = NUCL-EX 0502021;%%
%%CITATION = NUCL-EX 0106015;%%
%%CITATION = NUCL-EX 0204008;%%
%%CITATION = NUCL-EX 0309004;%%
%
\bibitem{KamLAND}
K.~Eguchi {\it et al.}  [KamLAND Collaboration],
{Phys.\ Rev.\ Lett.}  {\bf 90} (2003) 021802;
%%CITATION = HEP-EX 0212021;%%
%
T.~Araki {\it et al.}, hep-ex/0406035.
%%CITATION = HEP-EX 0212021;%%
%%CITATION = HEP-EX 0406035;%%
%
\bibitem{K2K}
E.~Aliu {\it et al.} [K2K Collaboration], hep-ex/0411038.
%%CITATION = HEP-EX 0411038;%%
%
\bibitem{STPNu04} S.~T.~Petcov, {Nucl.\ Phys.\ B (Proc. Suppl.)}
{\bf 143} (2005) 159.
%%CITATION = HEP-PH 0412410;%%
%
\bibitem{BPont57} B.~Pontecorvo, {Zh.\ Eksp.\ Teor.\ Fiz.\ (JETP)}
  {\bf 33} (1957) 549; {\it ibid.} {\bf 34} (1958) 247; {\it ibid.}
  {\bf 53} (1967) 1717;
%
 Z.~Maki, M.~Nakagawa and S.~Sakata, 
%%CITATION = PTPKA,28,870;%%
{Prog.\ Theor.\ Phys.} {\bf 28} (1962) 870.
%%CITATION = SPHJA,6,429;%%
%%CITATION = SPHJA,7,172;%%
%%CITATION = SPHJA,26,984;%%
%%CITATION = PTPKA,28,870;%%

\bibitem{MoscowH3Mainz}
V.~Lobashev {\it et al.},  
%%CITATION = NUPHZ,91,280;%%
{Nucl.\ Phys.} {\bf A719}(2003) 153c;
%
K.~Eitel {\it et al.}, {Nucl.\ Phys.\ B (Proc. 
Suppl.)} {\bf 143} (2005) 197.
%%CITATION = NUPHZ,143,197;%%
%
\bibitem{WMAPnu} D.~N.~Spergel {\it et al.} [WMAP Collaboration],
{Astrophys.\ J.\ Suppl.} {\bf 148} (2003) 175.
%%CITATION = ASTRO-PH 0302209;%%
%
\bibitem{Hanne03} S.~Hannestad, astro-ph/0303076;
 O.~Elgaroy and O.~Lahav, astro-ph/0303089.
%%CITATION = ASTRO-PH 0303076;%%
%%CITATION = ASTRO-PH 0303089;%%
%
\bibitem{seesaw} P. Minkowski, 
{Phys.\ Lett.} {\bf B67} (1977) 421;
M.~Gell-Mann, P.~Ramond, 
and R.~Slansky in {\it Supergravity},
p. 315, edited by F. Nieuwenhuizen 
and D.~Friedman, North Holland, Amsterdam, 1979;
T.~Yanagida, Proc. of the 
{\it Workshop on Unified Theories and the Baryon
Number of the Universe}, edited by 
O.~Sawada and A.~Sugamoto, KEK, Japan 1979;
R.~N.~Mohapatra, G.~Senjanovi{\'c}, 
{Phys.\ Rev.\ Lett.} {\bf 44} (1980) 912.
%
\bibitem{Pont67} 
B.~Pontecorvo, {Zh.\ Eksp.\ Teor.\ Fiz.} {\bf 53} (1967) 1717;
%%CITATION = SPHJA,53,1717;%%
S.~M.~Bilenky and B.~Pontecorvo,
{Lett.\ Nuov.\ Cim.} {\bf 17} (1976) 569.
%%CITATION = SPHJA,53,1717;%%
%
\bibitem{LeptoG} M.~Fukugita and T.~Yanagida, 
{Phys.\ Lett.} {\bf B174} (1986) 45.
%%CITATION = PHLTA,B174,45;%%
%
%
\bibitem{BiPet87} S.~M.~Bilenky and S.~T.~Petcov, 
{Rev.\ Mod.\ Phys.}  {\bf 59} (1987) 671.
%%CITATION = RMPHA,59,671;%%
%
\bibitem{mega}
M.~L.~Brooks {\it et al.} [MEGA Collaboration], 
{Phys.\ Rev.\ Lett.} {\bf 83} (1999) 1521.
%%CITATION = HEP-EX 9905013;%%
%
\bibitem{PDG04} S.~Eidelman {\it et al.} [Particle Data Group],
{Phys.\ Lett. } {\bf B592} (2004) 1.
%
\bibitem{BaBar05} B.~Aubert {\it et al.}  [BABAR Collaboration], 
hep-ex/0502032; hep-ex/0508012;
K.~Hayasaka {\it et al.} [BELLE Collaboration],
Phys.\ Lett.\ B {\bf 613} (2005) 20.
%%CITATION = HEP-EX 0502032;%%
%%CITATION = HEP-EX 0508012;%%
%%CITATION = HEP-EX 0501068;%% 
%
\bibitem{Kuno99} Y.~Kuno and Y.~Okada, {Rev.\ Mod.\ Phys.} 
{\bf 73} (2001) 151;
% \bibitem{Akeroyd:2004mj}
A.~G.~Akeroyd {\it et al.}  [SuperKEKB Physics Working Group],
%``Physics at super B factory,''
hep-ex/0406071.
%%CITATION = HEP-EX 0406071;%%
%%CITATION = RMPHA,73,151;%%
%
\bibitem{psi} L.~M.~Barkov {\it et al.}, the MEG Proposal (1999),
http://meg.psi.ch. 
%
\bibitem{BorzMas86}
F.~Borzumati and A.~Masiero, {Phys.\ Rev.\ Lett.} {\bf 57} (1986) 961.
%
\bibitem{GMFB} R.~Barbieri, S.~Ferrara and C.~Savoy,
{Phys.\ Lett. } {\bf B119} (1982) 343;
L.~Hall, J.~Lykken and S.~Weinberg, 
{Phys.\ Rev.} {\bf D27} (1983) 2359.
%
\bibitem{Hisano96} J.~Hisano {\it et al.},
{Phys.\ Lett.} {\bf B357} (1995) 579; {Phys.\ Rev.} {\bf D53} (1996) 2442;
J.~Hisano and D.~Nomura, {Phys. Rev.} {\bf D59} (1999) 116005.
%
\bibitem{Iba01} J.~A.~Casas and A.~Ibarra,
{Nucl.\ Phys.} {\bf B618} (2001) 171.
%%CITATION = HEP-PH 0103065;%%
%
\bibitem{JohnE} J.~Ellis {\it et al.},
{Nucl.\ Phys.} {\bf B621} (2002) 208; {Phys.\ Rev.} {\bf D66} (2002) 115013;
J.~Ellis and M.~Raidal, {Nucl.\ Phys.} {\bf B643} (2002) 229;
J.~Ellis, M.~Raidal and T.~Yanagida, 
{Phys.\ Lett.} {\bf B546} (2002) 228.
%%CITATION = HEP-PH 0109125;%%
%%CITATION = HEP-PH 0206110;%%
%%CITATION = HEP-PH 0206174;%%
%%CITATION = HEP-PH 0206300;%%
%
\bibitem{Saclay0105} 
S.~Lavignac, I.~Masina and C.~A.~Savoy,
{Phys.\ Lett.} {\bf B520} (2001) 269;
%
A.~Kageyama {\it et al.}, 
{Phys.\ Rev.} {\bf D65} (2002) 096010;
{Phys.\ Lett.} {\bf B527} (2002) 206;
%
F.~Deppisch {\it et al.}, 
{Eur.\ Phys.\ J.} {\bf C28} (2003) 365;
% 
X.-J. Bi,
{Eur. \ Phys. \ J.} {\bf C27} (2003) 399;
T.~Blazek and S.~F.~King,
{Nucl. Phys.} {\bf B662} (2003) 359;
%
B.~Dutta and R.~N.~Mohapatra,
{Phys.\ Rev.} {\bf D68} (2003) 056006;
%
J.~I.~Illana and M.~Masip,
{Eur. Phys. J.} {\bf C35} (2004) 365;
%
A.~Masiero, S.~K.~Vempati and O.~Vives,
{New J.\ Phys.} {\bf 6} (2004) 202;
%
M.~Bando {\it et al.}, hep-ph/0405071;
%
I.~Masina and C.~A.~Savoy, 
{Phys.\ Rev.} {\bf D71} (2005) 093003;
%
K.~S.~Babu, J.~C.~Pati and P.~Rastogi, hep-ph/0502152;
%
P.~Paradisi, hep-ph/0505046.
%%CITATION = HEP-PH 0106245;%%
%%CITATION = HEP-PH 0112359;%%
%%CITATION = HEP-PH 0110283;%%
%%CITATION = HEP-PH 0211236;%%
%%CITATION = HEP-PH 0206122;%%
%%CITATION = HEP-PH 0211368;%%
%%CITATION = HEP-PH 0305059;%%
%%CITATION = HEP-PH 0310257;%%
%%CITATION = HEP-PH 0407325;%%
%%CITATION = HEP-PH 0405071;%%
%%CITATION = HEP-PH 0501166;%%
%%CITATION = HEP-PH 0502152;%%
%%CITATION = HEP-PH 0505046;%%
%
\bibitem{PPY03} S.~Pascoli, S.~T.~Petcov and C.~E.~Yaguna,
{Phys.\ Lett.} {\bf B564} (2003) 241.
%%CITATION = HEP-PH 0301095;%%
%
\bibitem{PPR3} S.~Pascoli, S.~T.~Petcov and W.~Rodejohann,
{Phys.\ Rev.} {\bf D68} (2003) 093007;
W.~Rodejohann,
%``Hierarchical matrices in the see-saw mechanism, 
% large neutrino mixing  and leptogenesis,''
Eur.\ Phys.\ J.\ C {\bf 32} (2004) 235.
%  [arXiv:hep-ph/0311142].
%%CITATION = HEP-PH 0302054;%%
%%CITATION = HEP-PH 0311142;%%
%
\bibitem{PPTY03} S.~T.~Petcov {\it et al.},
{Nucl.\ Phys.} {\bf B676} (2004) 453.
%%CITATION = HEP-PH 0306195;%%
%
\bibitem{Eiichi05} 
S.~Kanemura {\it et al.}, hep-ph/0501228; hep-ph/0507264.
%
\bibitem{PShinYasu05} 
S.~T.~Petcov, T.~Shindou and Y.~Takanishi,
  %``Majorana CP-violating phases, RG running of neutrino mixing parameters and
  %charged lepton flavour violating decays,''
hep-ph/0508243.
%%CITATION = HEP-PH 0508243;%%
%
\bibitem{SP76} S.~T.~Petcov, 
{Sov.\ J.\ Nucl.\ Phys.} {\bf 25} (1977) 340.
%
\bibitem{BPP77} S.~M.~Bilenky, S.~T.~Petcov and B.~Pontecorvo, {Phys.\ 
    Lett.}  {\bf B67} (1977) 309; T.~P.~Cheng and L.-F.~Li, {Phys.\ 
    Rev.\ Lett.} {\bf 45} (1980) 1908.
%
\bibitem{LGBDiBP05} W.~Buchm{\"u}ller, P.~Di~Bari and M.~Pl{\"u}macher,
{New J. Phys.} {\bf 6} (2004) 105.
%%CITATION = HEP-PH 0406014;%%
%
\bibitem{CERN04} G.~F.~Giudice {\it et al.},
% A. Notari, M. Raidal, A. Riotto, A. Strumia
{Nucl. Phys.} {\bf B685} (2004) 89.
%%CITATION = HEP-PH 0310123;%%
%
\bibitem{BHP80} S.~M.~Bilenky, J.~Hosek and S.~T.~Petcov,
              {Phys.\ Lett.} {\bf B94} (1980) 495.
%%CITATION = PHLTA,B94,495;%%
%
\bibitem{Lang87} P.~Langacker {\it et al.},
{Nucl.\ Phys.} {\bf B282} (1987) 589.
%%CITATION = NUPHA,B282,589;%%
%
\bibitem{APSbb0nu} C.~Aalseth {\it et al.}, hep-ph/0412300.
%%CITATION = HEP-PH 0412300;%%
%
\bibitem{BPP1} S.~M.~Bilenky, S.~Pascoli and S.~T.~Petcov,
              {Phys.\ Rev.} {\bf D64} (2001) 053010.
%%CITATION = HEP-PH 0102265;%%
%
\bibitem{STPFocusNu04} S.~T.~Petcov, New J.\ Phys. {\bf 6} (2004) 109
({\it http://stacks.iop.org/1367-2630/6/109});
Talk given at the Nobel Symposium (N 129) on Neutrino Physics,
August 19 - 24, 2004, Haga Slot, Enk{\"o}ping, Sweden,
hep-ph/0504110; 
S.~Pascoli, S.~T.~Petcov and L.~Wolfenstein,
{Phys.\ Lett.} {\bf B524} (2002) 319;
S.~Pascoli, S.~T.~Petcov and T.~Schwetz, hep-ph/0505226;
S.~Choubey and W.~Rodejohann, hep-ph/0506102.
%%CITATION = HEP-PH 0504110;%%
%%CITATION = HEP-PH 0110287;%%
%%CITATION = HEP-PH 0505226;%%
%%CITATION = HEP-PH 0506102;%%
%
\bibitem{RGrunU} S.~Antusch {\it et al.},
{Nucl.\ Phys.} {\bf B674} (2003) 401.
%%%CITATION = HEP-PH 0305273;%%
%
\bibitem{GUTM3} 
R.~N.~Mohapatra {\it et al.},
%(Part of the APS Neutrino Study), 
hep-ph/0510213;
G.~Altarelli and F.~Feruglio,
{New J. Phys.} {\bf 6} (2004) 106;
% hep-ph/0405048;
R.~Dermisek and S.~Raby, hep-ph/0507045;
C.~H.~Albright, {Phys. Rev.} {\bf D72} (2005) 013001.
%%CITATION = HEP-PH 0510213;%%
%%CITATION = HEP-PH 0405048;%%
%%CITATION = HEP-PH 0507045;%%
%%CITATION = HEP-PH 0502161;%%
%
\bibitem{LHCSUSY} A.~Airapetian {\it et al.} [ATLAS Collaboration],
Report CERN-LHCC-99-15;
S.~Abdullin {\it et al.}  [CMS Collaboration],
%``Discovery potential for supersymmetry in CMS,''
J.\ Phys.\ G {\bf 28} (2002) 469.
%[arXiv:hep-ph/9806366]
%%CITATION = HEP-PH 9806366;%%
%

\bibitem{SchValle80D81} M.~Doi {\it et al.},
{Phys.\ Lett.} {\bf B102} (1981) 323;
J.~Schechter and J.~W.~F.~Valle, 
{Phys.\ Rev.} {\bf D22} (1980) 2227;
J.~Bernabeu and P.~Pascual,
{Nucl.\ Phys.} {\bf B228} (1983) 21.
%%CITATION = NUPHA,B228,21;%%
%%CITATION = PHRVA,D22,2227;%%
%%CITATION = PHLTA,B102,323;%%
%
\bibitem{CHOOZPV} 
M.~Apollonio {\it et al.}, {Phys.\ Lett.} {\bf B466} (1999) 415;
%%CITATION = HEP-EX 9907037;%%
%
F.~Boehm {\it et al.}, 
{Phys.\ Rev.\ Lett.}  {\bf 84} (2000)  3764. 
%%CITATION = HEP-EX 9912050;%%
%
\bibitem{BCGPRKL2} A.~Bandyopadhyay {\it et al.},
{Phys.\ Lett.} {\bf B608} (2005) 115;
A.~Bandyopadhyay {\it et al.}, 2005, unpublished; see 
also  A.~Bandyopadhyay  {\it et al.},
{Phys. Lett.} {\bf B583} (2004) 134.
%%CITATION = HEP-PH 0406328;%%
%%CITATION = HEP-PH 0309174;%%
%
\bibitem{3nuGlobal} 
J.~N.~Bahcall, M.~C.~Gonzalez-Garcia and C.~Pe\~na-Garay, 
{JHEP} {\bf 0408} (2004) 016.
%%CITATION = HEP-PH 0405172;%%               
%
\bibitem{PPSNO2bb} S.~Pascoli and S.~T.~Petcov,
{Phys.\ Lett.} {\bf B544} (2002) 239; {\it ibid.} 
{\bf B580} (2004) 280.
%%CITATION = HEP-PH 0205022;%%
%%CITATION = HEP-PH 0310003;%%
%
%
\bibitem{FGY03} P.~H.~Frampton, S.~L.~Glashow and T.~Yanagida,
{Phys. Lett.} {\bf B548} (2002) 119.
%%CITATION = HEP-PH 0208157;%%
%
\bibitem{daviba02} 
S.~Davidson and A.~Ibarra, {Phys. Lett.} {\bf B535} (2002) 25.
%%CITATION = HEP-PH 0202239;%%
%
\bibitem{BGKP96} S.~M.~Bilenky {\it et al.}, 
{Phys.\ Rev.} {\bf D56} (1996) 4432.
%%CITATION = HEP-PH 9604364;%%

\end{thebibliography}
\end{document}